\documentclass[%
preprintnumbers,
pre,
amsmath,amssymb,
aps,
]{revtex4-2}
\usepackage{graphicx}
\usepackage[hyperfootnotes=false]{hyperref}
\usepackage{wasysym}
\usepackage{subcaption}
\usepackage{color, colortbl}
\usepackage{bm}% bold math
\definecolor{Gray}{gray}{0.9}
\definecolor{LightCyan}{rgb}{0.88,1,1}
\usepackage[first=0,last=9]{lcg}

\usepackage[dvipsnames]{xcolor}

\newcommand{\beginsupplement}{%
       \setcounter{table}{0}
       \renewcommand{\thetable}{S\arabic{table}}%
       \setcounter{figure}{0}
       \renewcommand{\thefigure}{S\arabic{figure}}%
    }

\newcommand{\R}{\mathcal{R}}

\date{\today}
\begin{document}

\author{Michael Neuder}
\author{Elizabeth Bradley}%
 \altaffiliation[Also at]{
 Santa Fe Institute, Santa Fe, New Mexico, 87501
 }
\affiliation{%
 Department of Computer Science\\
 University of Colorado\\
 Boulder, Colorado, 80309
 }%

\author{Edward Dlugokencky}
\affiliation{
 National Oceanic and Atmospheric Administration\\
 Boulder, Colorado, 80305
}%

\author{James W. C. White}
\affiliation{%
 Institute of Arctic and Alpine Research\\
  University of Colorado\\
 Boulder, Colorado, 80309
}%

\author{Joshua Garland}
\email{To whom correspondence should be addressed; E-mail:  joshua@santafe.edu.}
\affiliation{%
 Santa Fe Institute\\ 
 Santa Fe, New Mexico, 87501
}%

\preprint{APS/Physical Review E Submission}

\title{Detection of Local Mixing in Time-Series Data Using Permutation Entropy}

\begin{abstract}
  While it is tempting in experimental practice to seek as high a
  data rate as possible, oversampling can become an issue if one
  takes measurements too densely.  These effects can take many
  forms, some of which are easy to detect: e.g., when the data
  sequence contains multiple copies of the same measured value.  In
  other situations, as when there is mixing---in the measurement
  apparatus and/or the system itself---oversampling effects can be
  harder to detect.  We propose a novel, model-free technique to
  detect local mixing in time series using an information-theoretic
  technique called permutation entropy.  By varying the temporal
  resolution of the calculation and analyzing the patterns in the
  results, we can determine whether the data are mixed locally, and
  on what scale.  This can be used by practitioners to choose
  appropriate lower bounds on scales at which to measure or report
  data.  After validating this technique on several synthetic
  examples, we demonstrate its effectiveness on data from a
  chemistry experiment, methane records
  from Mauna Loa, and an Antarctic ice core.
\end{abstract}

\maketitle

\section{Introduction}
In many fields of science, data are measured in gas or fluid states
where local mixing occurs: in measurement chambers and laboratory
piping, for instance.  This can artificially interchange information
between neighboring data points and thereby obfuscate the results.
Similar effects can also be at work in the system under study: e.g.,
diffusion of isotopes in an ice sheet, which mixes climate data from
successive years.  In the face of these challenges, choosing an
appropriate interval at which to measure, analyze, or report the data
is often an imperfect balance between time, money, laboratory
capabilities, and scientific need.  Frequently, educated guesses are
the only way to make these critical choices.

In this paper, we describe a novel technique to detect local mixing in
a data set: not only whether or not it is present, but also (if so) on
what scale.  This gives practitioners a way to know precisely where to
draw the line, in terms of sampling the system and reporting the data.
Critically, our approach is model free, providing results without any
need for domain-specific knowledge.  The technique is based on a
method called permutation entropy, which provides an estimate of the
rate at which new information is produced in a data sequence: in
effect, a measure of predictability.  By varying the ``stride'' of
this calculation, we can detect whether or not the data bear the scars
of mixing.  The underlying idea is as follows: for most time series,
predictability tends to decrease as one extends the horizon,
so permutation entropy will generally increase with the stride of the
calculation because the data points involved span wider and wider
temporal ranges.  If successive data points are measured on a smaller
scale than the mixing scales that are inherent in the data, though,
each measurement is essentially a single draw from a local
distribution composed of the data points in its local neighborhood. 
By definition, this added randomness will raise the entropy rate.
Reversal of the normal pattern of the relationship between the stride
of the calculation and the permutation entropy values, then, is an
effective way to detect local mixing.  Used in tandem with judicious
bin averaging, this method also allows one to detect the scale of
those mixing effects and adjust one's procedures accordingly.

In Section~\ref{sec:methods}, we describe the techniques involved in
our approach, beginning with some background on permutation entropy
and describing the metric that we have developed to capture the
effects described in the previous paragraph.  We validate our
technique using two synthetic examples (Sections~\ref{sec:lorenz}
and~\ref{sec:mg}) and explore its utility in the context of three
real-world time-series data sets: one from a chemistry experiment
involving gas mixtures (Section~\ref{sec:gas}), a second from an
Antarctic ice core (Section~\ref{sec:wdc}), and a third from the Mauna
Loa methane records (Section~\ref{sec:mauna-loa}).  We discuss these
results and their implications in Section~\ref{sec:discussion} and
conclude, with some thoughts about broader applications and future
directions, in Section~\ref{sec:concl}.

\section{Materials and Methods}\label{sec:methods}

\subsection{Permutation Entropy}\label{sec:pe}

Permutation entropy \cite{bandt2002permutation} is a method for
estimating the Shannon entropy rate\cite{amigo2010permutation} of an
arbitrary time-series data set.  The calculation focuses on the local
relationships among a sequence of points by mapping their values to an
ordering of the same length.  The three-point sequence $[7,3,11]$, for
instance, would map to the ordering $[1,0,2]$ because $3 < 7 < 11$.
The statistics of these ordinal sequences or {\sl permutations} are
calculated as follows.  Let $|\cdot|$ denote set cardinality and
\texttt{Ord} define a function that calculates the ordering of a given
sequence of length $\ell$, as in the $\ell=3$ example above.
Additionally, define a time series as the sequence $(x_i)_{i \in
  I_N}$, where the index set is $I_N = \{k \in \mathbb{N}\setminus\{0\} \;|\; k \leq
N \}$.
Then the probability of the appearance in the time
series of $\ell$-sequences of data points whose values map to a
specific ordering $\pi$ can be estimated by
\begin{align} \label{eq:PEwotau}
    \mathrm{Pr}[\pi] &= \frac{|\{n \in I_{N-\ell+1} \;|\;
      \texttt{Ord}(x_n, x_{n+1}, ..., x_{n+\ell-1}) = \pi \}|}{N - \ell +
      1}.
\end{align}

\begin{figure}
 \centering \includegraphics[width=0.8\textwidth]{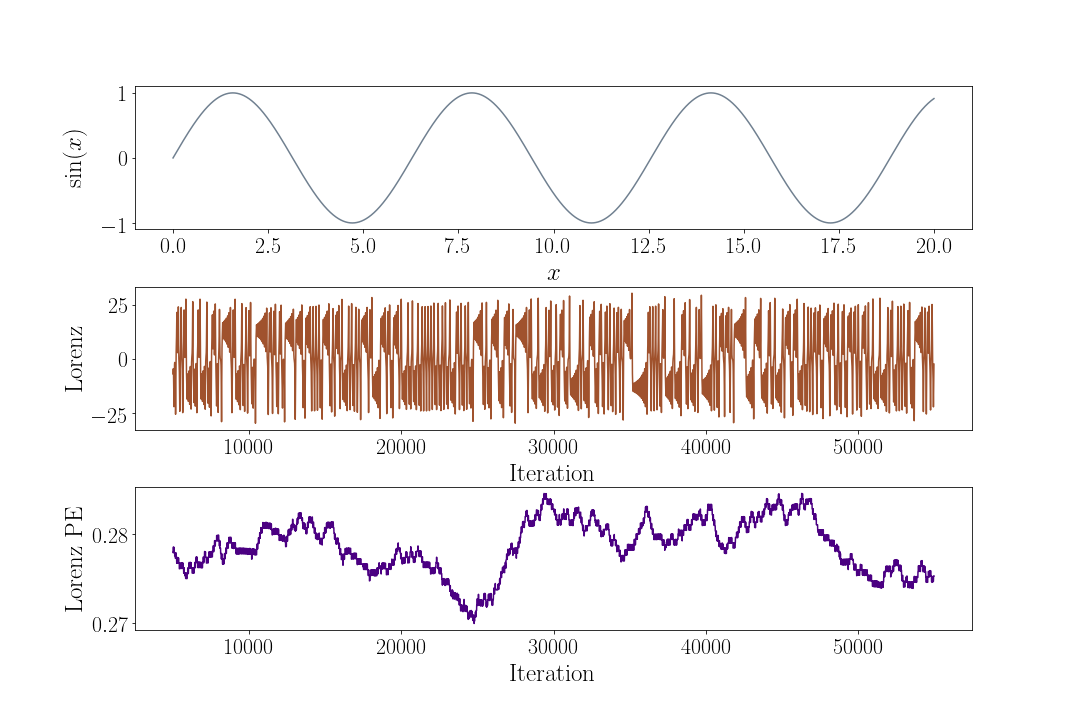}
\caption{\textbf{Demonstration signals} Top: sinusoidal signal.
  Middle: a short segment of a chaotic signal from the Lorenz system.
  Bottom: permutation entropy of the Lorenz time series in the middle
  panel.  % Full traces appear in the Supplementary Materials.
% , calculated with a   word length of five (i.e., $\ell=4$) in a 
% sliding window of 5000 points (i.e., $w=5000$).
}
\label{fig:PE-ex}
\end{figure}

By calculating the probability of each of the $\ell!$ orderings across
the whole time series, one constructs the permutation entropy or $PE$:
\begin{align}
    PE &= - \frac{1}{\log \ell !} \sum_{\pi} \mathrm{Pr}[\pi] \log \mathrm{Pr}[\pi].
\end{align}
This is an estimate of the average rate at which new information
appears in a time series per observation\cite{bandt2002permutation}.
With the $1/\log\ell!$ normalization, its values range from 0 to 1.
If $PE$ is low, the observations, on average, contain a significant
amount of information about the past.  This is the case for the sine
wave in the top panel of Figure~\ref{fig:PE-ex} where $PE\approx
0.012$.  
If $PE$ is high, on the other hand, most of the information in each
observation is new.  As such, this has been shown to correlate with
predictability
\cite{pennekamp2019intrinsic,garland2014model,scarpino2019predictability}.
This is the case for the time series in the middle panel of
Figure~\ref{fig:PE-ex} from the canonical Lorenz system, described at
more length later in this paper, where $PE \approx 0.28$, reflecting
the limited predictability that is associated with deterministic
chaos.\footnote{The integration step size used to generate the time
  series also plays a role here; if it is very small, successive
  points are highly correlated and the $PE$ will be lower.}  If the
signal were {\sl wholly} random, the $PE$ would be 1.  The value of
the $PE$ is a function of the subsequence length $\ell$, of course;
please see
\cite{myers2020automatic,pennekamp2019intrinsic,riedl2013practical}
for more discussion of this parameter and its implications.

As defined above, $PE$ captures the permutation entropy of an entire
data set in the form of a single value.  One can also perform this
calculation in a sliding window across a time series to increase the
temporal resolution of the results.  This is a good idea, for example,
 if the dynamics of the system at hand are complex or nonstationary.
In this variant of the calculation, each $PE$ value captures the
statistics of the orderings in a window around the associated time
point.

As shown in the bottom panel of Figure~\ref{fig:PE-ex}, this windowed
analysis brings out subtle changes in the predictability of the Lorenz
time series in the middle panel as it moves through different regimes
in its behavior.  However, it does introduce another free parameter
into the calculation; see
\cite{cao2004detecting,garland2019information} for discussion of the
associated choices and issues.  

All of the calculations reported in this paper use $\ell=4$ and
$w=5000$, the former chosen via a careful persistence analysis that
seeks stabilization of the results with variations in this parameter,
and the latter as dictated by the formulas in
\cite{garland2014model,garland2018anomaly,riedl2013practical,Amigo2008}.
In a window of this length, each permutation of length four will have
approximately 200 opportunities to appear at least once, which should
be enough to rule out any forbidden ordinals\cite{Amigo2008}.

\subsection{Permutation entropy at varying temporal resolutions}

The examples and equations in the previous section assume that one
always works with ordinal permutations that are constructed from the
values of {\sl successive} points in a time series.  It can also be
useful to coarsen the grain of this procedure using $\tau$-separated
points instead, thereby changing the temporal resolution of the
calculation.  This variant of the $PE$ technique, which was introduced
in \cite{cao2004detecting}, requires a few modifications to the first
step in the calculations.  Specifically, Equation~(\ref{eq:PEwotau})
becomes
\begin{align} \label{eq:PEwtau} \mathrm{Pr}[\pi] &=
  \frac{|\{n \in I_{N-(\ell+1)\tau} \;|\; \texttt{Ord}(x_n, x_{n+\tau},
    ..., x_{n+(\ell-1)\tau}) = \pi \}|}{N-(\ell+1)\tau}
\end{align}
where $\tau$ defines the spacing between the points in the time series
that are used to construct the $\ell$-element ordinal permutations.  With
$\tau=2$, for instance, the first permutation constructed from the
sequence $[1,4,6,2,5,3]$ would be $[0,2,1]$ because $1 < 5 < 6$.  

\begin{figure}
    \centering
\includegraphics[width=0.6\textwidth]{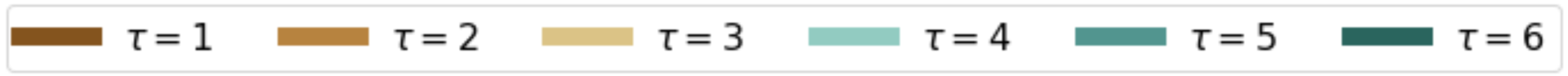}

\vspace*{1mm}

\begin{subfigure}{0.32\textwidth}
    \centering 
\includegraphics[width=1\textwidth]{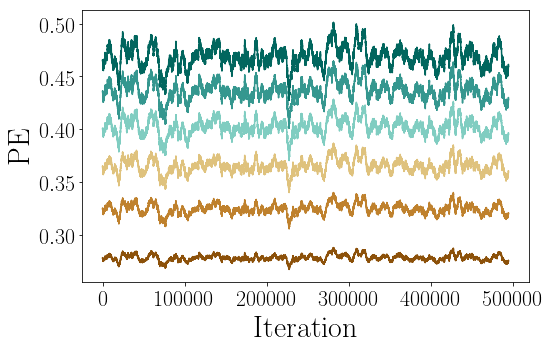}
\caption{}
\end{subfigure}
\begin{subfigure}{0.32\textwidth}
    \centering 
\includegraphics[width=1\textwidth]{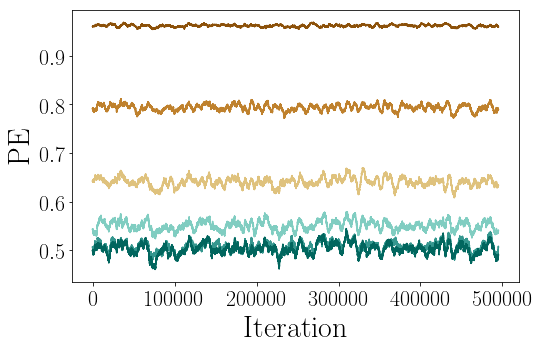}
\caption{}
\end{subfigure}
\begin{subfigure}{0.32\textwidth}
    \centering 
\includegraphics[width=1\textwidth]{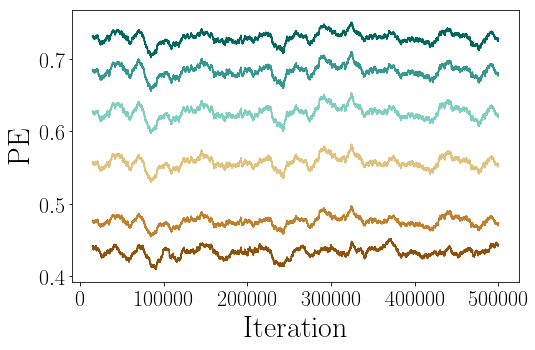}
\caption{}
\end{subfigure}

\caption{\textbf{Lorenz permutation entropy.} (a) Permutation entropy
  ($PE$) of the full version of the Lorenz signal from the middle
  panel of Figure~\ref{fig:PE-ex}, calculated with a range of values
  of $\tau$, the spacing between data points used to construct the
  permutations.  (The trace in the bottom panel of Figure
  \ref{fig:PE-ex} is a short segment of the $\tau=1$ trace here.)
  Note the pattern of increasing $PE$ with $\tau$: i.e., the $\tau=a$
  trace is higher than the $\tau=b$ trace for $a>b$. (b) If local
  mixing effects are artificially added to the same signal,
the ordering of the $PE$ traces is reversed.  (c) A simple bin
averaging operation removes the local-mixing effects, restoring the
normal order of the traces.}
\label{fig:lorenz-varying-tau}
\end{figure}

The spacing parameter $\tau$ is the focus of the work reported here,
and the central point of leverage for our technique.  Because this
parameter controls the ``stride'' of the calculation, changing its
value allows one to understand the information dynamics of the system
at different resolutions.  This reveals some interesting patterns:
persistence of features in $PE$ results across a range of $\tau$
values, for instance, indicates an effect in the underlying signal
that spans multiple time scales.  And in a wide range of data sets, we
have observed that \emph{increasing the $\tau$ value generally raises
  the $PE$ curves}.  These data sets include a number of classic
chaotic systems, two of which are used as examples later in this
paper, as well as transient deterministic dynamics (e.g., the examples
used in \cite{cao2004detecting}), various paleoclimate records (e.g.,
\cite{garland2019information,garland2016first}), various one-second
and one-minute financial price datasets, computer performance data
from the experiments reported in
\cite{myktowicz2009,garland2011predicting}, and experimental data from
a driven damped pendulum, which are available at \cite{liz-pend-data}.
Figure~\ref{fig:lorenz-varying-tau}(a) demonstrates this in the context
of the Lorenz signal from Figure~\ref{fig:PE-ex}.

\begin{figure}
    \centering \includegraphics[width=0.7\textwidth]{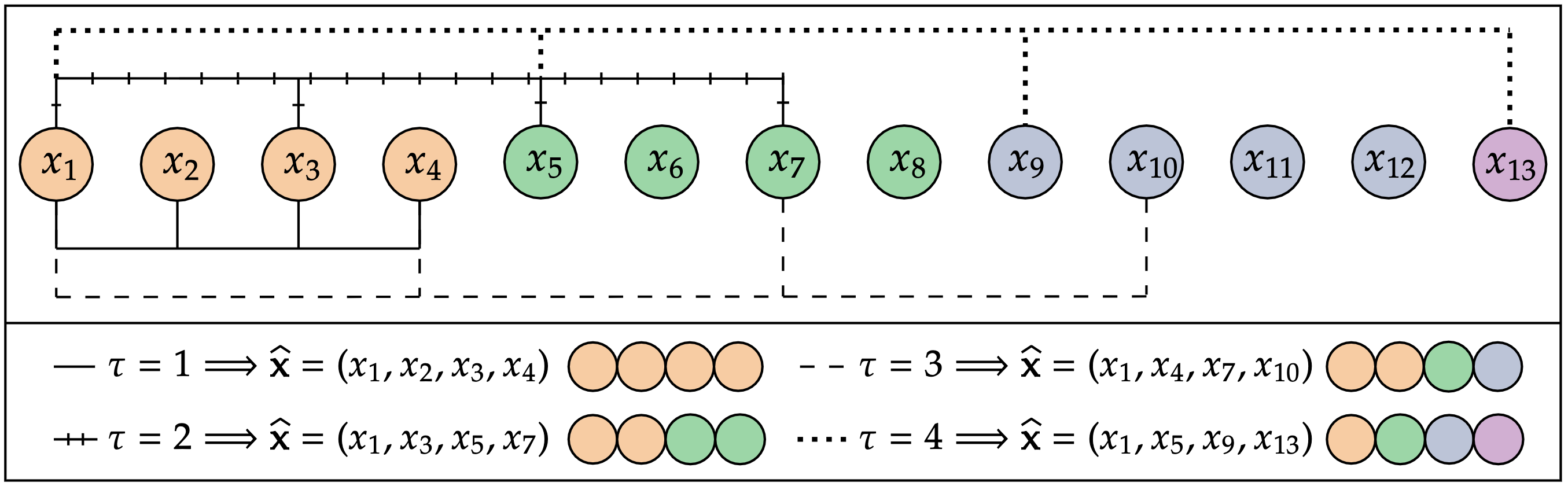}
    \caption{\textbf{Mixing and permutation entropy.} If the values of
      neighboring points are not independent and the $\tau$ value is
      small, then the points used to construct each permutation are
      effectively drawn from a single local distribution, schematized
      here with different colors.  In this case, even though the
      underlying data may be deterministic, the relationships of these
      draws will be random and the permutation entropy will be
      artificially high.  Higher $\tau$ values widen the span of the
      calculation beyond the spread of the local distributions,
      thereby mitigating this effect.}
    \label{fig:cause}
\end{figure}

Note that the values of the permutation entropy increase monotonically
with $\tau$.  This simply reflects decreasing predictability over the
longer time span sampled by each permutation.  When there is local
mixing in a time series, though, that reasoning no longer applies.
Rather, because the data points used to construct individual
permutations are measured on a smaller scale than the mixing scales
that are inherent in the data, each element of those permutations is
essentially a single draw from a local distribution composed of the
data points in the neighborhood of each individual measurement.  By
definition, this added randomness will raise the $PE$.  As $\tau$
increases, though, the points used to generate the orderings will
spread out across those local distributions.  When that point spacing
exceeds the local distribution width---i.e., the largest mixing scale
at work in the data---the extrinsic $PE$
increase~\cite{pennekamp2019intrinsic} caused by that mixing will be
reduced and eventually eliminated.  The schematic in
Figure~\ref{fig:cause} illustrates the mechanism underlying this
effect.

The takeaway here is that, when local mixing is at work in a data set,
its $PE$ will not increase monotonically with $\tau$.  Rather, the
vertical ordering of the $PE$ curves in a plot like the ones in
Figure~\ref{fig:lorenz-varying-tau} will be reversed, with the
low-$\tau$ traces at the top and the high-$\tau$ traces at the bottom:
the opposite of what we expect for a deterministic system.  This
reversal is not only a clear indicator of a specific issue in the
data.  With some additional experimentation, as described in
Section~\ref{sec:datasamptech}, this effect can actually bring out the
scale of the mixing.  First, though, we need a formal metric for
assessing the relationships between permutation entropy curves
calculated with different values of $\tau$.

\subsection{Reversal Metric}
\label{sec:reversal-metric}

The purely monotone nature of the vertical orderings of the different
$PE$ traces in Figure~\ref{fig:lorenz-varying-tau} persists across the
entire span of the time series, but that is not always the case; for
some data sets, the traces will touch, or even cross, at different
points in the time series.  In order to assess the overall form of
these patterns, what we want is a measure of how close this vertical
ordering is to a purely monotone-increasing or monotone-decreasing
pattern at different points in the time series.  To quantify this
formally, we develop the following metric.  Let $\tau_{min}$ and
$\tau_{max}$ denote the smallest and largest values of $\tau$ used in
the $PE$ calculation.  Additionally, let $\mathbf{v}_i$ denote the vector of monotone-increasing 
values of $\tau$:
\begin{align}
    \mathbf{v}_i &= 
    \begin{bmatrix}
        \tau_{min} & \tau_{min+1} & ... & \tau_{max}
    \end{bmatrix}^{T} \nonumber 
\end{align} 
If $\tau_{min}=1$ and $\tau_{max}=6$, for instance, this vector
would be:
\begin{align}
    \mathbf{v}_i &= 
    \begin{bmatrix}
        1 & 2 & 3 & 4 & 5 & 6
    \end{bmatrix}^{T} \nonumber 
\end{align} 
\label{page:vectors}
\noindent We construct a {\sl focal $\tau$-sequence vector},
$\mathbf{v}$, that captures the ordering of the values of $PE$ at a
given time $n$.  (For the example in
Figure~\ref{fig:lorenz-varying-tau}, $\mathbf{v}=[1 \; 2 \; 3 \; 4 \;
  5 \; 6]^T \; \forall \; n$.)  We define the reversal metric $\R$ as
the distance between this focal $\tau$-sequence vector and the purely
monotone-increasing vector $\mathbf{v}_i$:
\begin{equation} \label{eq:Reqn}
    \R(\mathbf{v}) = || \mathbf{v}
    - \mathbf{v}_i ||/\lambda
\end{equation}
\noindent where $\mathbf{v}\in\mathcal{S}_{\tau_{\max}}$, the set of
permutations of order $\tau_{max}$, and $||\cdot||$ denotes the L1
norm.  The scaling term $\lambda$, which is defined as
\begin{equation}
\lambda=\max_\mathbf{y}||\mathbf{y}-\mathbf{v_i}|| \nonumber
\end{equation}
with $\mathbf{y}\in\mathcal{S}_{\tau_{max}}$, normalizes the value of
$\R$ to run from 0 to 1, where 0 indicates perfect monotone-increasing
order of the different $PE$ curves with $\tau$.  This is the case for
every time point in Figure~\ref{fig:lorenz-varying-tau}(a).  $\R=1$,
on the other hand, indicates perfect monotone-decreasing order---as in
Figure~\ref{fig:lorenz-varying-tau}(b), where local-mixing effects
have been added to the Lorenz signal via the procedure described in
the following section.  One can average these pointwise values of $\R$
across the time series, or over subsequences of it, in order to assess
the overall reversal pattern in the corresponding span.  In the
examples in Figure~\ref{fig:lorenz-varying-tau}(a) and~(b), that
average value $\overline{\R}$ is identical for all subsequences of the
data.  If the reversal patterns vary across the time series, though,
that will not be the case---as will be shown in
Section~\ref{sec:mauna-loa}.

The way of measuring distance that is encoded in
Equation~(\ref{eq:Reqn}), which is a normalized version of Spearman's
footrule \cite{diaconis1977spearman,kendall1948rank}, is more appropriate in our application than
the Kendall-$\tau$ distance \cite{abdi2007kendall,kendall1948rank}---which is
essentially the number of pairs on which the relative order of two
permutations disagrees---because a reversal of, say, the $\tau=1$ and
$\tau=6$ traces is more meaningful than a reversal of the $\tau=3$ and
$\tau=4$ traces because of the time scales involved.
Other choices are of course possible for the form of this metric, as
discussed in Section~\ref{sec:discussion}.

 \subsection{Data Manipulation Techniques} \label{sec:datasamptech}

To validate our conjectures about the effect of mixing on the $\tau$
orderings of $PE$ traces, we perform a number of tests using synthetic
time-series data sets $\mathbf{x} = (x_n)_{n\in I_N}$ from well-known
dynamical systems, manipulating them in order to create ansatzes that
replicate the effects of mixing on those data: specifically, how it
causes the interchange of information within some mixing window that
includes $k$ points before and after each measurement.  To do this, we
create a local normal distribution at each point $\mathbf{x}_n$ whose
mean and variance are calculated from the points in that mixing window
around that point.  To produce the corresponding point of the ansatz,
$\hat{\mathbf{x}}_n$, we make a random draw from that distribution:
%   That is, each point in the mixed series is drawn from the 
%   distribution
\[\hat{\mathbf{x}}_n {\sim}
    \textrm{Normal}(\mu_n, \sigma_n^2) \] where $\mu_n$ and $\sigma_n$
    are the mean and standard deviation of the points of $\mathbf{x}$
    in the window:
\begin{align}
[\mathbf{x}_{n - k},\dots,\mathbf{x}_n,\dots,\mathbf{x}_{n + k}]
\nonumber
\end{align}
This is a comparatively simple model of the effects of mixing on a
series of data points; it does not capture long-range effects, for
instance.  But for a wide range of experimental situations, this is a
useful approximation.

In the following section, we also make use of binned averaging, where
one uses groups of $j$ consecutive points of a time series
$\mathbf{y}$ to create the elements of a new series
$\overline{\mathbf{y}}$:
\begin{align}
     \overline{\mathbf{y}}_n &= \left(\frac{y_{(n-1)j+1} + y_{(n-1)j + 2} + ... + 
y_{nj}}{j}\right)_{n \in I_{N/j}} \label{eq:bin}
 \end{align}
Note that this is not a {\sl rolling} bin average; rather, the first
point of $\overline{\mathbf{y}}$ is the average of the first $j$ points of
$\mathbf{y}$, the second is the average of
$[\mathbf{y}_{j+1}...\mathbf{y}_{2j}]$, and so on.  This is intended
to mimic what happens in experimental practice when $j$ successive
points in a raw data set are averaged together---in the system, in the
post processing, or in the laboratory apparatus.  In the analyses that
follow, this technique serves two purposes:
as an extra validation step in the synthetic examples and as a
diagnostic of mixing scales in the real-world data sets.

\section{Results}\label{sec:results}

\begin{figure}
    \centering
    \includegraphics[width=0.8\textwidth]{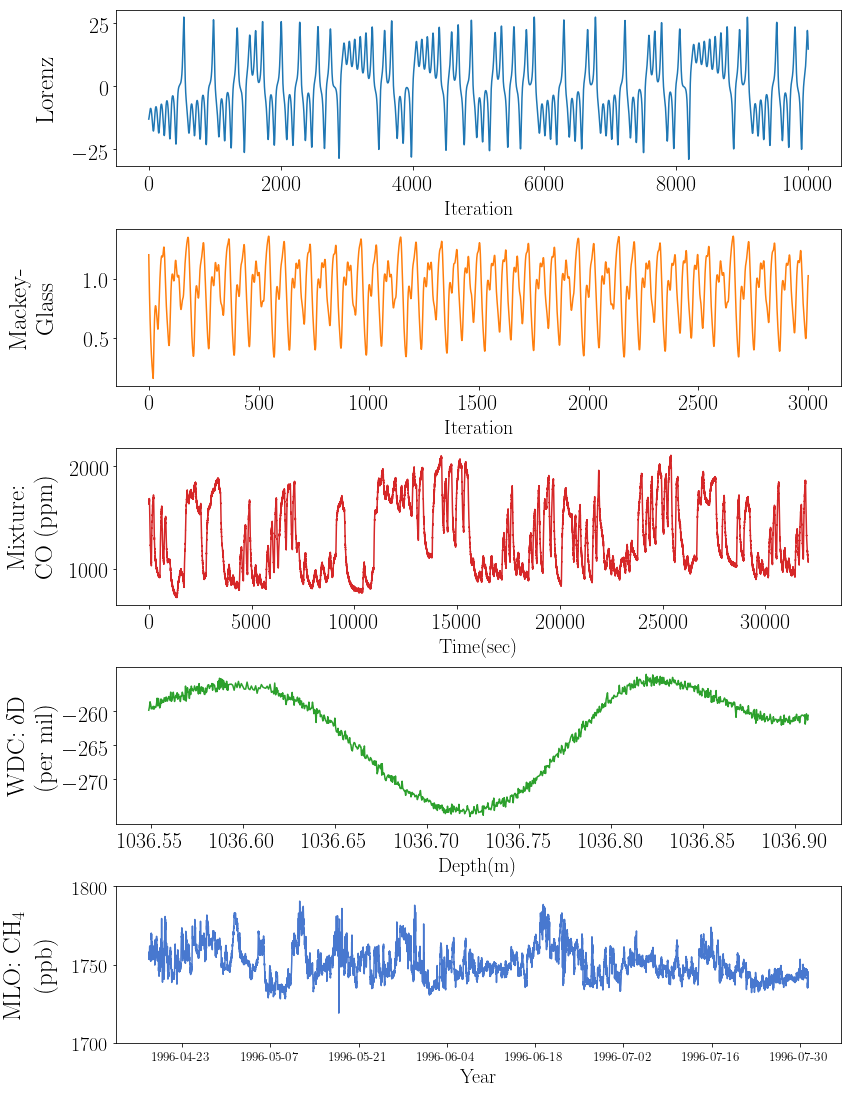}
    \caption{\textbf{Data sample.} A sample from each of the four
      time-series data sets used as examples in this paper: the
      classic Lorenz and Mackey-Glass systems from nonlinear dynamics,
      chemical sensor data from a gas-mixing experiment, water-isotope
      data from an Antarctic ice core (WDC), and atmospheric methane
      records from Mauna Loa (MLO).}
    \label{fig:data_sample}
\end{figure}

As validation and demonstration cases for our mixing-detection
technique, we use two synthetic examples---classic systems from the
field of nonlinear dynamics---and three real-world time-series data
sets.  The synthetic-data examples allow us to manipulate the time
series, as described in Section~\ref{sec:datasamptech}, in order to
validate our reasoning about mixing and $\tau$ reversal.  The
experimental data sets---chemical sensor measurements from a gas
mixing experiment, water-isotope data from an Antarctic ice core,
measured by a spectrometer, and atmospheric methane from the Global
Monitoring Laboratory observatory on Mauna Loa, Hawai'i---provide a
first view into the utility of this technique in real-world settings.
Figure \ref{fig:data_sample} shows a data sample from each of these
time series.

\subsection{Lorenz system}\label{sec:lorenz}

The Lorenz system \cite{lorenz1963deterministic} is defined by the
following ordinary differential equations:
\begin{align}
    \Dot{x} &= a (y - x) \nonumber \\
    \Dot{y} &= x (r - z) - y  \nonumber \\
    \Dot{z} &= xy - bz \nonumber 
\end{align}
To create a synthetic data set from this system, we choose $a=16, \;
b=4,$ and $r=45$---parameter values that are known to produce chaotic
behavior---and integrate the ODEs using a $4^{th}$ order Runge-Kutta
method for 500,000 steps with a step size of $h=0.005$, starting from
the initial condition $[x_0, y_0, z_0] = [-13, -12, 52]$.  The $x$
coordinate of this trajectory---the first 55,000 points of which are
shown in the top trace in Figure~\ref{fig:data_sample}---is the
baseline time series $\mathbf{x}$ for the set of tests reported in
this section.

Using the process described in Section~\ref{sec:datasamptech}, we
manipulate this high-resolution trajectory in order to replicate the
effects of local mixing in real experiments.  The details are as
follows.  At each point $\mathbf{x}_n$ in the original time series, we
calculate the mean and variance of the points in a window of width
six, centered on that point (i.e.,
$[\mathbf{x}_{n-3}...\mathbf{x}_{n+3}]$) and then make a random draw
from a normal distribution with those parameters to obtain the
corresponding point of a new time series $\hat{\mathbf{x}}_n$.  We
choose a mixing window of width six for this first experiment because
$\ell=3$ gives us points from many distributions for each $\tau$.

As discussed previously, this $\hat{\mathbf{x}}$ signal---whose $PE$
is shown Figure~\ref{fig:lorenz-varying-tau}(b)---is an ansatz that is
specifically designed to let us explore the effects of local mixing in
a controlled experiment.  As expected, the monotone-increasing
vertical order that is apparent in the $PE$ traces calculated from the
original time series is reversed by the local-mixing operation.  The
corresponding value of the reversal metric $\overline{\R}$, computed across
the span of the $PE$ traces in Figure~\ref{fig:lorenz-varying-tau}(b),
is 1, reflecting the perfect monotone-decreasing order of those traces
with $\tau$.  (This would also be the case for $\overline{\R}$ computed
across any {\sl subsequence} of those traces, since the ordering is reversed
everywhere: i.e., $\R=1 \; \forall \; n$.)  Note that the $PE$ values are
higher than in Figure~\ref{fig:lorenz-varying-tau}(a).  This simply
reflects the additional randomness introduced by the local mixing
operation.  Careful examination shows that that effect decreases with
$\tau$: the $\tau=6$ trace in Figure~\ref{fig:lorenz-varying-tau}(b),
for instance, has roughly the same mean value as in
Figure~\ref{fig:lorenz-varying-tau}(a), but the movement in the
smaller-$\tau$ traces is comparatively larger.  This is due to the
mitigation of the mixing effects that occurs with larger $\tau$
values.

\begin{figure}
    \centering
    \includegraphics[width=0.45\textwidth]{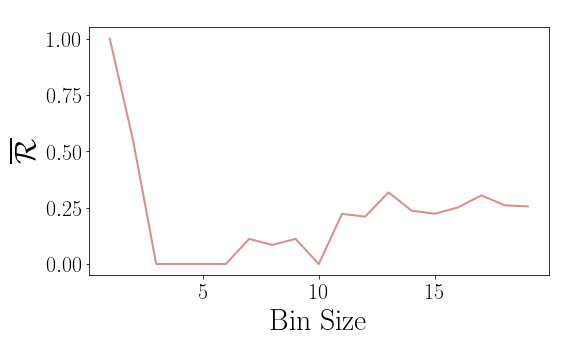}
    \includegraphics[width=0.45\textwidth]{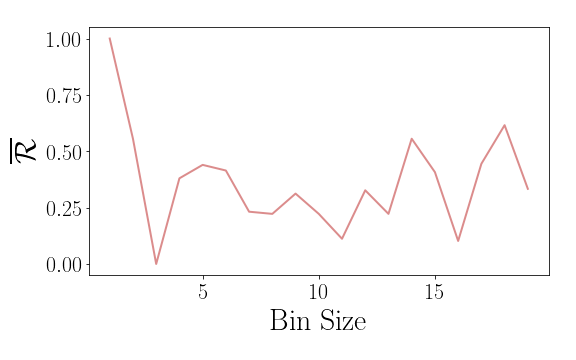}

\vspace*{-2mm}
\textbf{(a)} \hspace*{2.9truein} \textbf{(b)}

    \caption{\textbf{Bin size and mixing scales.} Plots of $\overline{\R}$
      versus bin size for the (a) Lorenz and (b) Mackey-Glass
      examples.}
    \label{fig:Rbar-vs-binsize}
\end{figure}

Our original conjecture was that one could effectively remove the
added local mixing effects in $\hat{\mathbf{x}}$ by taking a binned
average of that signal using the technique described in
Section~\ref{sec:datasamptech}, with a bin size $j$ roughly equal to
the size of the mixing window used in the creation of that ansatz.
This conjecture can be easily tested using the ansatz signal, as
binning at that scale should ``undo'' the $\tau$ reversal produced by
that synthetic mixing manipulation, thereby restoring the normal
ordering.\footnote{For systems with delays, such as those modeled with
  delay-differential equations, a wider binning window may be required
  because of the associated temporal propagation of information due to
  the delay term.}
We propose the following heuristic for choosing the value of $j$ to
remove the local-mixing effects: bin-average the time series using
varying bin sizes $j$, calculating $\overline{\R}$ across the span of the
time series at each one, and choose the first bin size where
$\overline{\R}$ goes to zero; if $\overline{\R}$ never reaches zero, choose the
first minimum of this curve.  Figure~\ref{fig:Rbar-vs-binsize}(a)
shows such a plot for the Lorenz ansatz.
As the bin size increases, the curves start reordering themselves,
with $\overline{\R}$ reaching zero at $j=3$.  Beyond that, the ordering
shifts because of the complicated relationships between the different
timescales of the analysis ($\tau$, $\ell$, the mixing window, and the
bin size).  The $PE$ traces in Figure~\ref{fig:lorenz-varying-tau}(c),
computed from the Lorenz ansatz $\hat{\mathbf{x}}$ binned with
$j=3$,
are in clear monotone-increasing order with $\tau$, consistent with
the $\overline{\R}$ value of 0.

These results confirm that the binning process---a data-manipulation
step that manipulates the ansatz to eliminate the simulated effects of
local mixing---does indeed restore the normal ordering.  Note that the
bin size indicated by our heuristic is smaller than the local mixing
window that was used to create the ansatz, so our original conjecture
was off by a scale factor.  This is discussed further below.  Note,
too, that the binning operation does not restore {\sl the exact form}
of the original traces in Figure~\ref{fig:lorenz-varying-tau}(a), just
their ordering.  The $PE$ values in
Figure~\ref{fig:lorenz-varying-tau}(c) are higher because of the
residual effects of the added mixing.  The difference in the
smoothness of the traces in these two images is partially due to the
altered window size---from the reduction in the number of points in
the $\overline{\mathbf{x}}$ signal---as well as the difference in the
scales of the images.

This series of experiments validates the conjecture that, when local
mixing occurs in a data set, high sampling rates can lead to reversal
of the normal ordering of permutation entropy with changing $\tau$.
{\em This observation is a potentially useful way to detect the
  presence of local mixing in experimental data.}  The bin size that
restores the correct ordering is an indicator of the scale of the
effect, and that size can be chosen from a curve like the one in
Figure~\ref{fig:Rbar-vs-binsize}(a).  Our experiments show that this
claim holds for different mixing windows, as well as for different $w$
and $\ell$ values.  In all cases, that bin size is smaller than, but
on the same order as, the mixing window used in the creation of the
ansatz.  The implications of this, for experimental practice, are that
this operation {\em effectively identifies the order of magnitude of
  the mixing scale in the data, though not its specific value.}

These results are encouraging, but this is only one system, and a
low-dimensional one at that.  In the following section, we replicate
these steps with a more-complicated dynamical system.

\subsection{Mackey-Glass System}
\label{sec:mg}

The Mackey-Glass system \cite{Mackey287} is described by a
delay-differential equation of the following form:
\vspace*{-3mm}
\begin{align} \label{eq:MG}
    \Dot{x} &= \beta \frac{x(t-t_0)}{1+x(t-t_0)^q} - \gamma x, \quad \gamma, \beta, q > 0.
\end{align}
The delay here makes this system effectively infinite
dimensional, even though there is only a single state variable.  Like
the Lorenz system, Mackey-Glass is known to exhibit deterministic
chaos for some values of the parameters $\gamma, \beta, t_0 \textrm{
  and } q$.  Using $\beta=0.2, \gamma=0.1, t_0=17, \textrm{ and }
q=10$, we integrate Equation~(\ref{eq:MG}) with the same Runge-Kutta
solver for $1.5 \times 10^6$ steps from an initial state of $x = 1.2$
using a step size of $0.1$ to obtain our test data $\mathbf{x}$.  The
first 3000 points of this time series are shown as the second trace in
Figure~\ref{fig:data_sample}.

\begin{figure}
    \centering \includegraphics[width=0.6\textwidth]{img/flipLegend.png}
\vspace*{1mm}

    \centering
    \includegraphics[width=0.3\textwidth]{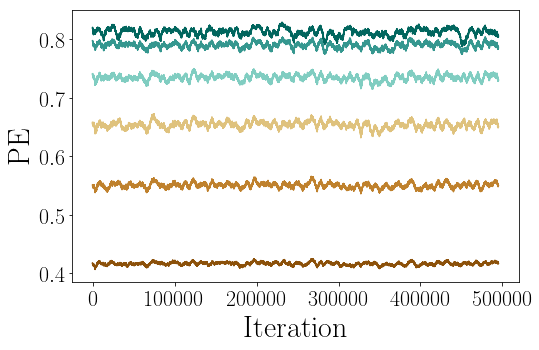}
\vspace*{-2mm}
    \includegraphics[width=0.3\textwidth]{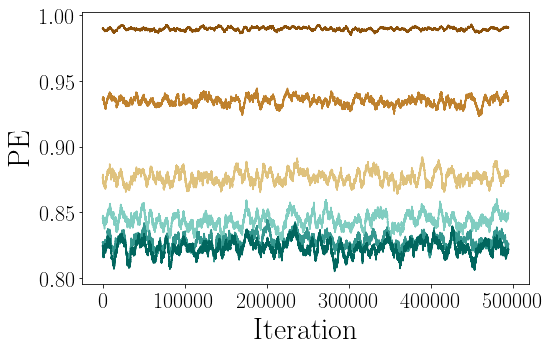}
\vspace*{-2mm}
    \includegraphics[width=0.3\textwidth]{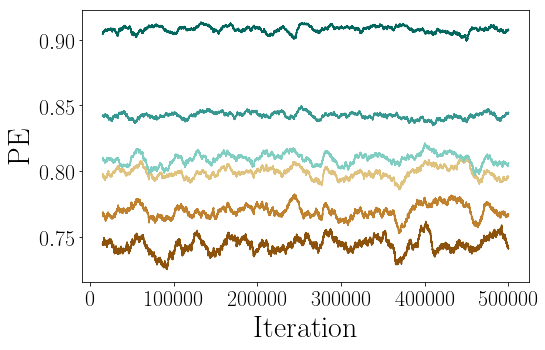}
\vspace*{3mm}

\textbf{(a)} \hspace*{1.5truein} \textbf{(b)} \hspace*{1.5truein} \textbf{(c)}
    \caption{\textbf{Reversal results in the Mackey-Glass system.}
      $PE$ traces from the Mackey-Glass system with
      $1\le\tau\le6$ for (a) the original time series (b) with added
      local-mixing effects (c) after a bin-averaging operation was
      applied to remove those effects.  
}
    \label{fig:mg-results}
\end{figure}

The procedure for exploring $\tau$ reversal in this example is exactly
the same as in the Lorenz system in the previous section.
Figure~\ref{fig:mg-results} shows $PE$ traces of the ${\mathbf{x}}$,
$\hat{\mathbf{x}}$ and $\overline{\mathbf{x}}$ signals.
\noindent
Figure~\ref{fig:mg-results}(a) demonstrates normal,
monotone-increasing order, with $\R=0$ at all time points and a
corresponding value of $\overline{\R}=0$ computed across the span of the
time series.  For the locally mixed ansatz $\hat{\mathbf{x}}$ in
Figure~\ref{fig:mg-results}(b), which was created with a mixing window
of eight, the order is reversed and $\overline{\R}=1$.  As before, bin
averaging with a bin size chosen at the first minimum of the
$\overline{\R}$ versus $j$ curve for this signal
(Figure~\ref{fig:Rbar-vs-binsize}(b)) completely restores the
monotone-increasing order of the $PE$ traces, as shown in
Figure~\ref{fig:mg-results}(c).  Again, the form of these results
persists for different mixing windows, as well as for different values
of the parameters of the $PE$ calculations.  And as in the Lorenz
examples, the bin size needed to fully restore the normal ordering is
always smaller than, but on the same order as, the mixing window used
in the creation of the ansatz.  In other words, this value provides a
useful litmus test for the {\sl scale} of that effect, even if it does
not indicate the exact value.

These results reaffirm the overall relationship between local mixing
effects and $\tau$ reversal.  Next, we move to data from physical
experiments.

\subsection{Gas Mixture Data}
\label{sec:gas}

As a first real-world demonstration case for these techniques, we use
a data set from the UCI Machine Learning repository \cite{uciMLrepo}
that was produced during a chemistry experiment involving gas mixing.
The values in this time series were measured by a sensor measuring
carbon monoxide and ethylene mixtures in air under changing
concentration conditions \cite{fonollosa2015reservoir}.  We focus on
the $9^{th}$ sensor in the array, a Firago TGS-2600 instrument that
reports particle concentrations in PPM.
% Mike picked #9 because it had large values
These data have an extremely high background noise level, which raises
all the $PE$ curves to the top of their ranges and 
scrambles their order, so we first perform a filtering step.  The
third plot in Figure~\ref{fig:data_sample} shows 30,000 data points
from the resulting time series, which is spaced at 0.25 sec.  See the
Supplementary Material for the full trace.

\begin{figure}
    \centering
\includegraphics[width=0.6\textwidth]{img/flipLegend.png}

\vspace*{1mm}

\begin{subfigure}{0.4\textwidth}
    \centering 
\includegraphics[width=1\textwidth]{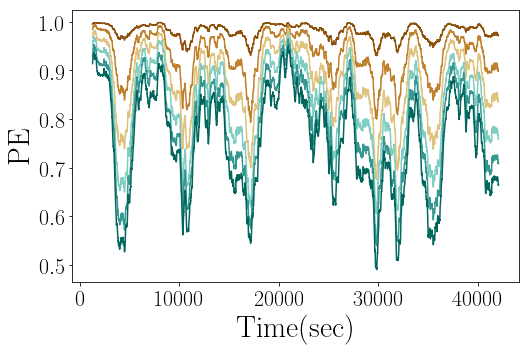}
\caption{Gas Mixture -- Raw}
\end{subfigure}
\begin{subfigure}{0.4\textwidth}
    \centering 
\includegraphics[width=1\textwidth]{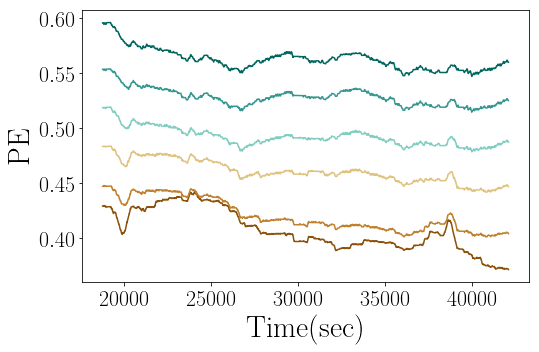}
\caption{Gas Mixture -- Binned with $j=15$}
\end{subfigure}

\begin{subfigure}{0.4\textwidth}
    \centering 
\includegraphics[width=1\textwidth]{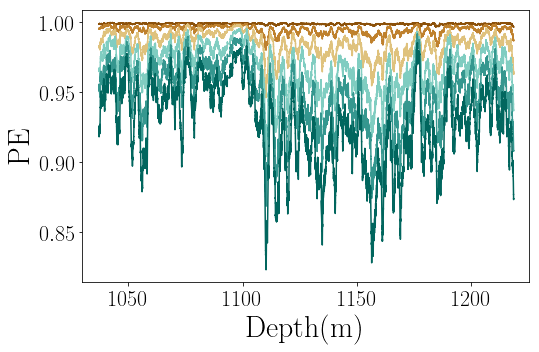}
\caption{WDC -- Raw}
\end{subfigure}
\begin{subfigure}{0.4\textwidth}
    \centering 
\includegraphics[width=1\textwidth]{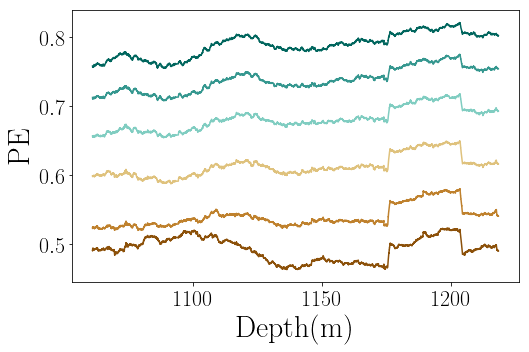}
\caption{WDC -- Binned with $j=15$}
\end{subfigure}

\begin{subfigure}{0.4\textwidth}
    \centering 
\includegraphics[width=1\textwidth]{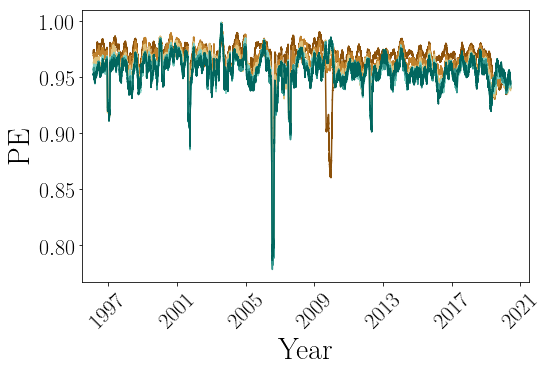}
\caption{Mauna Loa -- Raw}
\end{subfigure}
\begin{subfigure}{0.4\textwidth}
    \centering 
\includegraphics[width=1\textwidth]{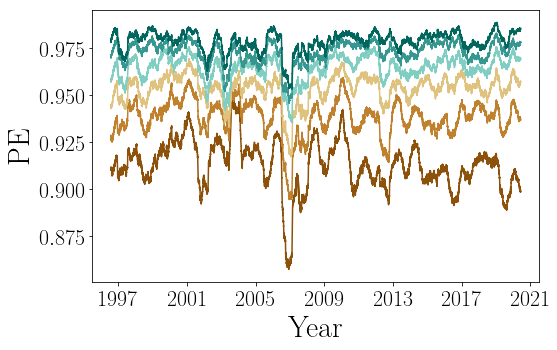}
\caption{Mauna Loa -- Binned with $j=4$}
\end{subfigure}
\caption{\textbf{Reversal results in experimental data sets.}  From
  top to bottom: $PE$ traces for the gas mixture, WAIS Divide, and
  Mauna Loa data.  In all cases, the images in the left and right
  columns show the calculations for the raw and bin-averaged data,
  respectively.  The traces in the latter are temporally offset by the
  width of the binning window times the width of the $PE$ calculation
  window.  (This offset is essentially invisible in the WDC data
  because of the number of data points in the time series.)}
\label{fig:exp-results}
\end{figure}

$PE$ traces calculated from this time series demonstrate reversed
$\tau$ ordering across the entire span of the data set, as shown in
the plot in Figure~\ref{fig:exp-results}(a), with $\overline{\R}=1$.
This suggests that local mixing was at work during the collection of
these data: in the chamber, perhaps, or the compartment in the sensor.
To ascertain whether this was indeed the case, we perform a bin
averaging step on the data---as in the synthetic examples in the
previous sections---and observe how that operation affects the
ordering of the $PE$ curves.  And as in those previous examples, we
can repeat this analysis using varying bin sizes in order to estimate
the scale of the mixing effects.

\begin{figure}[h]
    \centering
\includegraphics[width=0.33\textwidth]{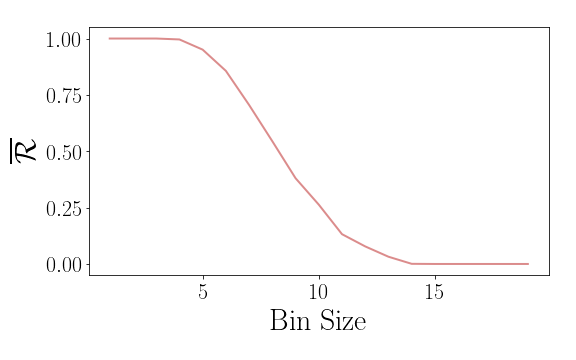}
\hspace*{-4mm}
\includegraphics[width=0.33\textwidth]{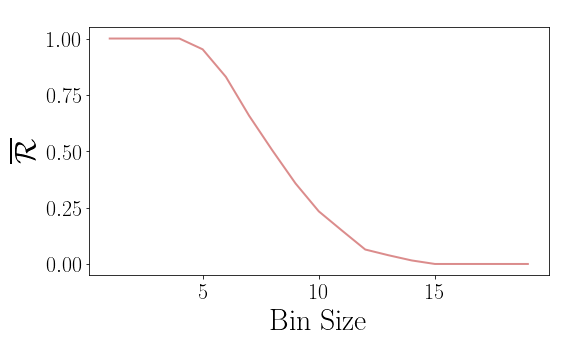}
\hspace*{-4mm}
\includegraphics[width=0.33\textwidth]{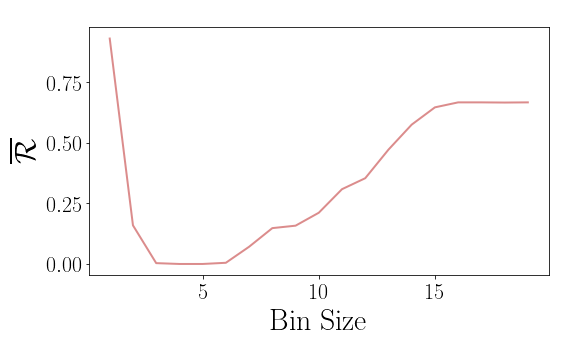}

\textbf{(a)} \hspace*{1.5truein} \textbf{(b)} \hspace*{1.5truein} \textbf{(c)}

\caption{\textbf{$\overline{\R}$ as a function of bin size} (a) gas mixture
  (b) WAIS Divide and (c) Mauna Loa Methane}
\label{fig:rbar-binsize}
\end{figure}

In this example, the results of that procedure are quite clear, though
the shape of the $\overline{\R}$ versus bin size curve is somewhat
different than in our synthetic examples; see
Figure~\ref{fig:rbar-binsize}(a).
As before, the $PE$ traces move towards normal ordering as the bin
size increases, but $\overline{\R}$ then reaches a broad minimum that
begins at $j=15$ and extends to $j=54$ (not shown).\footnote{We do
  not have enough data to extend the calculation beyond $j=60$.}
Figure~\ref{fig:exp-results}(b) shows that the $PE$ traces for data
bin-averaged with $j=15$---i.e., 3.75 seconds---are in fully normal
order.  {\em These results not only confirm that local mixing is at
  work in the raw data, but also suggest a scale for those effects.}
This, in turn, has implications for the resolution at which these data
should be reported.

\subsection{Antarctic Ice-Core Data}
\label{sec:wdc}

Water isotopes in ice cores, which are viewed as climate proxies
\cite{Dansgaard-64-Tellus}, can be measured at very high resolution in
modern continuous flow analysis (CFA) equipment
\cite{jones2017improved}.  During this process, the ice is
continuously melted and then piped through tubing into an optical
spectrometer that measures the ratios of heavy to normal variants of
oxygen and hydrogen isotopes in the ice.  The CFA apparatus in the
Stable Isotope Laboratory at the Institute for Arctic and Alpine
Research at the University of Colorado can perform these measurements
at sub-millimeter resolution.  Basic fluid dynamics makes mixing
inevitable in the different stages of this analysis pipeline;
see~\cite{jones2017improved} for discussion of the associated issues.
Diffusion causes these isotopes to mix in the ice sheet as well,
before the samples reach the laboratory, so there are at least two
potential sources of mixing in these data---perhaps with different
scales.

The West Antarctic Ice Sheet Divide Core (WDC) is 3300m long, so a
data set gathered from this core with sub-millimeter spacing contains
many million points.  Here, we focus on a short segment of that
record: the ice from 1035.4 m--1368.2 m, which captures climate
information from roughly 4500-6500 years
ago~\cite{garland2018anomaly,WDCReSample}.  The fourth trace in
Figure~\ref{fig:data_sample} shows a small segment of the hydrogen
isotope ($\delta \text{D}$) data from this core; see the Supplementary
Materials for the full trace.  These data are reported in delta
($\delta$) notation relative to the baseline Vienna Standard Mean
Ocean Water (VSMOW) and normalized to the Standard Light Antarctic
Water (SLAP, $\delta \text{D}= -428.0$ \permil) scale.  The $\delta$
value is determined by $\delta=1000(\rho_{sample}/\rho_{VSMOW} - 1)$, where
$\rho$ is the isotopic ratio D/H (i.e., $^2\text{H}/^1\text{H}$).  Please
see~\cite{garland2019information} for more details about these data.
The sinusoidal pattern is a result of seasonal variability in
temperature.

As is customary in this field, the raw, high-resolution data---0.3 mm
spacing, in this data set---undergo a number of post-processing steps:
removal of outliers, interpolation to fill gaps where data are
missing, etc.  The final
step in the laboratory's post-processing procedure involves a binned
average of the data, with the bin scale explicitly chosen to reduce
any mixing effects.  In this case, the Stable Isotope lab used a bin
size of 0.5 cm (i.e., 17 points at an 0.3 mm spacing) to construct the
published version of this data, which is available at
\cite{WDCReSample}.

This provides a unique opportunity for our work: using permutation
entropy, we can not only assess whether mixing was indeed at work in
the raw data, but also potentially identify the associated scales, and
even validate our results against expert judgment about those scales.
To that end, we calculate $PE$ and $\overline{\R}$ for the raw 0.3 mm
spaced data.  Similarly to the gas mixture data in
Section~\ref{sec:gas}, the higher-$\tau$ $PE$ traces are below the
lower-$\tau$ traces across the full span of the data set; see
Figure~\ref{fig:exp-results}(c).  The corresponding $\overline{\R}=1$ value
reflects this consistent reversal of the normal monotone-increasing
order across the entire time series.  As in the gas-mixture example,
this suggests that local mixing effects are present in those data.
Note that $\tau$-reversal does not identify the {\sl mechanics} of the
effect, or its source.  In this example, there are many potential
culprits: the laboratory apparatus and the physics of the ice sheet,
among other things.  See Section~\ref{sec:discussion} for more
discussion.

Note that we have moved to speaking of spatial scales and resolutions
here---depths in the core---rather than temporal scales.  Over this
span of the WDC, there is a linear relationship between depth and age,
with 0.5 cm of ice translating to 1/34th of a year of climate record,
so these depthwise data can be viewed
%% for the purposes of this tau reversal analysis
as time-series data.  This would not be the case if we were using a
longer span of the core---particularly one that extended to its deeper
levels, where the nonlinearity of the age-depth relationship more
strongly affects the temporal spacing of the data points.  

To explore the mixing scales in this ice-core record, we again use bin
averaging, computing $\overline{\R}$ from the raw 0.3 mm data, binned at a
range of bin sizes.  The shape of this curve, which is shown in
Figure~\ref{fig:rbar-binsize}(b), is similar to the one for the
gas-mixture data, reaching a broad minimum at $j=15$ with
$\overline{\R}=0$.  That is, bin-averaging the raw data using $j=15$
restores the normal order of the $PE$ traces; this is clearly visible
in Figure~\ref{fig:exp-results}(d).  As before, this is an indication
of the scale of the mixing effects that are present in the data.  In
this case, where we have expert judgment as a point of comparison,
that scale can be corroborated.  Recall that the laboratory used a bin
size of 17 to construct the data to be reported on the U.S. Antarctic
Program Data Center website, basing that choice on expert knowledge
about the CFA apparatus.
% and including a $2\times$ safety margin.  
{\em In other words, our technique not only matches expert reasoning
  about the data, but does so in a manner that is completely
  model-free, requiring no knowledge about the system that produced
  the data or the apparatus that was used to measure it.}

This analysis brings out the effects of {\sl any} local mixing
process, of course---not just laboratory effects.  Since the isotopes
mix via diffusion in the ice sheet, there are also potential
scientific implications of these results, as discussed further in
Section~\ref{sec:discussion}.

\subsection{Mauna Loa Methane Data}
\label{sec:mauna-loa}

NOAA’s Global Monitoring Laboratory has been measuring atmospheric
composition from Mauna Loa Observatory in Hawaii (altitude = 3397 m)
at high temporal resolution for decades.  Air is pumped continuously
at $\approx 6 L/min$ from two inlets 40 m above the surface to a
manifold inside the laboratory where $\approx 100 mL/min$ of the flows
is fed to a calibrated analyzer that measures the abundance of
atmospheric CH4 at fixed intervals \cite{dlugokencky95}.  Over the
period on which we focus in this paper, two well-calibrated analysis
methods were used.  In both cases, the analyzer alternates between the
two inlets.  From 1996 to 9 April 2019, measurements were by gas
chromatography with flame ionization detection (GC/FID) at 15 minute
intervals, reported in ppb.\footnote{$10^-9$ mol $CH_4$ per mol dry
  air.}  For each measurement, a 10 mL (at STP) slug of air is
injected into the GC/FID system every 15 minutes.  The response of the
detector is compared with bracketing measurements of standard to
quantify $CH_4$.  From 11 April 2019 to present, a Cavity Ring-Down
Spectrometer (CRDS) was used.  This instrument records nine 10-second
averages of $CH_4$ every five minutes and averages them into a single
``five-minute average'' value, also reported in ppb.  It is calibrated
with a suite of standards every two weeks relative to a reference
cylinder of natural air; this reference is measured once per hour
during normal monitoring to track drift in the analyzer.

Because $PE$ calculations require data with a constant temporal
spacing, we first downsampled the 5 min segment of the data to produce
an 826,633 point time series at an even 15 minute spacing, beginning
at 0007 on 1 January 1996 and ending at 2350 on 7 June 2020.  We then
scanned the data for missing or damaged values, replacing each of them
with the last good value. The fourth plot in
Figure~\ref{fig:data_sample} shows a 10,000-point
% 3.5 month
segment of this record beginning in March of 1996; a plot of the full
data set appears in the Supplementary Materials.
% 
% numpy.datetime64('1996-01-01T00:07:00.000000000'),
% numpy.datetime64('2020-06-07T23:50:00.000000000')
% 
% and this is 826633 points. It is 15 minute data all the way
% through. The five minute data I downsample to 15 min data so the time
% scales match up

The ordering of the $PE$ traces of this important data set, which are
shown in Figure~\ref{fig:exp-results}(e), is more complex than in the
previous two examples.  The overall pattern is reversed, but not
perfectly, with $\overline{\R}=0.907$, rather than 1.0, as in all of the
previous examples.  In other words, local mixing effects are almost
certainly in play here, but the indication is not as clear as in the
gas-mixture and ice-core examples, where $\R=1$ at every point in time
(and $\overline{\R}=1$ for the whole span).  Moreover, nonstationarity is
obviously an issue: viz., the downward spikes in $PE$ values,
especially at $\tau=1$ (e.g., in 2004), for instance, and the short
regions where the $\tau$-ordering of the traces changes (e.g., in 2010
and from 2019-on).  Some of these issues can be traced back to the
data-cleaning and quality-control processes; many of the dips, for
instance, correspond to regions where a large fraction of the data are
missing or damaged.  Replacing those points with the last good value
effectively creates a constant signal which, by definition, obeys
normal $\tau$ ordering.  This is a well-known challenge in the
permutation entropy literature: one that causes problems for all
data-cleaning strategies, including ours
\cite{mccullough2016counting,sakellariou2016counting,pennekamp2019intrinsic}.

As before, we can repeat the $\overline{\R}$ calculation for a range of bin
sizes in order to confirm the presence of mixing and estimate the
associated scale.  In this case, as Figure~\ref{fig:rbar-binsize}(c)
shows, the first minimum of $\overline{\R}$ falls at $j=4$.  The $\overline{\R}$
value at this point is 0, indicating that a binning operation at a
scale of one hour eliminates the local mixing effects.  
% Binned = 2.7640725845460705e-05
This is reflected in the ordering of the resulting traces, shown in
Figure~\ref{fig:exp-results}(f), which are in monotone-increasing
order with $\tau$.  {\em Again, our results are in good alignment with
  standard practice in this field, which prescribes a one-hour
  reporting cadence.}  As in the gas-mixture and ice-core examples in
the previous section, this has implications for both analysis and
reporting.  Unlike those examples, though, the curve in
Figure~\ref{fig:exp-results}(f) rises strongly for $j>5$.  This
behavior, which more closely resembles the two synthetic examples in
Figure~\ref{fig:Rbar-vs-binsize}, is likely due to the time scales
involved, in the systems and in the calculations---as well as to data
limitations, which precluded a broader span on the horizontal axes of
Figure~\ref{fig:exp-results}(d) and~(e).

\begin{figure}
    \centering 
\includegraphics[width=0.7\textwidth]{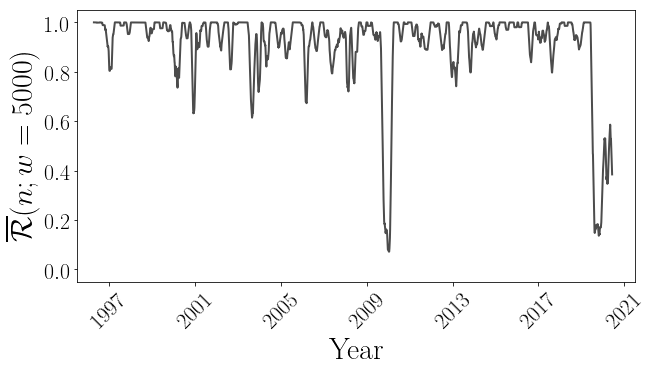}

\caption{Windowed $\overline{\R}$ Analysis for the Mauna Loa data}
\label{fig:windowed-rbar}
\end{figure}

When one suspects nonstationarities, it can be informative to perform
a more-focused analysis: specifically, to compute $\overline{\R}$ over {\sl
  windows} of the $PE$ traces, rather than over their entire temporal
extent.  Figure~\ref{fig:windowed-rbar} shows the results of this
calculation for the raw and binned MLO methane data that uses a window
size of 5000 points.
This plot brings out the nonstationarities of the data very clearly.
From 1996 through early 2019, $\overline{\R}$ oscillates between 1.0 and
0.7, with occasional drops to $\approx$0.6 and one even lower, in
2010.  These results confirm that the mixing effects in this segment
of the time series vary across the time series: the ordering of the
curves is largely reversed across most of the span of the data, but
the magnitude of that reversal varies with time.  
Many of these nonstationarities are due to the quality-control and
data-cleaning processes, as described above.  Between June 2009 and
May 2010, for instance, at the time of the obvious spike in the middle
of the curve, 5066 of the 31000 points were missing or damaged.
Standard interpolation strategies for filling these gaps---a necessary
step because $PE$ calculations require data that are evenly spaced in
time---employ smooth functions: lines, curves, etc.  Data spaced in
this manner are inherently predictable, so the $PE$ values in the
repaired spans will be low and the $\tau$ ordering will be monotone
increasing, hence the drop in $\overline{\R}$.  These effects are visible
in $PE$ traces themselves---Figure~\ref{fig:exp-results}(e) and the
magnified views that appear in the Supplementary Materials, which zoom
in on the time spans around 2000 and 2009---but they really stand out
in the windowed $\overline{\R}$ plot in Figure~\ref{fig:windowed-rbar}.
The same effects are also at work in many of the smaller dropouts in
$\overline{\R}$: e.g., a temporary 10X increase in analyzer noise that
occurred in June 2003, causing many of the points in this time span to
be flagged during the quality-control process as suspicious, and
therefore removed in our data-cleaning process.
% 
% The attached figure suggests that the change in mid-2003 was a
% degradation of GC precision. Each aliquot (slug) of air analyzed on
% the GC is bracketed by aliquots of standard, which track drift and
% calibrate the GC's output. By ignoring the air samples, I can look at
% the short-term noise of the analyzer, plotted here on the y-axis as
% the standard deviation of the normalized injections of standard gas
% from a cylinder. You can see about a factor of 10 increase in
% short-term noise in June, 2003. This is matched with the decrease in
% R-bar.

One of the most interesting features of Figure~\ref{fig:windowed-rbar}
is the precipitous drop in $\overline{\R}$ in the spring of 2019.  Recall
that the Observatory switched from a gas chromatograph to a cavity
ring-down spectrometer at 0155 on 11 April 2019, at the same time as
the shift in measurement cadence.
This newer instrument has a factor of 5 to 10 less short-term noise
than the gas chromatograph that it replaced.  {\em Our metric detects
  this quite effectively from the raw data, not only flagging the
  timing of the change, but indicating clearly that the introduction
  of this new instrument significantly reduced the mixing effects in
  the data.}  This provides additional corroboration of the efficacy
of our technique, and also has implications for the scientific
analysis, as described in the following section.

\section{Discussion}
\label{sec:discussion}

The synthetic-data experiments in Sections~\ref{sec:lorenz}
and~\ref{sec:mg}, which use ansatzes that are expressly designed to
mimic the effects of local mixing, confirm our conjectures about
$\tau$-reversal of $PE$ curves as an indicator of local mixing in a
data set.  In both cases, the $\overline{\R}$ metric shows that the
monotone-increasing order of the $PE$ traces for the original data is
reversed when information is artificially interchanged between
neighboring data points.  In both examples, a bin-averaging operation
on these artificially mixed time series restores the normal order---as
expected, since that operation compresses the distribution that is
effectively created by that mixing down to its mean.  The bin sizes
required to accomplish this are on the same scale as, but somewhat
smaller than, the width of the mixing window that we use to create the
ansatzes.  One can leverage this observation to estimate the mixing
scale: specifically, when the $\overline{\R}$ value of the $PE$ curves
reaches zero.  

Applied in tandem with judicious bin averaging, this reversal metric
has an even more powerful role in the examples in
Sections~\ref{sec:gas},~\ref{sec:wdc}, and~\ref{sec:mauna-loa}, which
involve data from real-world experiments.  In all of these cases
except a few spans of the Mauna Loa data, the $\tau$-reversal metric
suggests that the resolution of these data sets was actually high
enough that the raw data are essentially repeated draws from a
distribution composed from the data in a local window around each
point.  Permutation entropy brings this out quite effectively, if one
varies the $\tau$ parameter in the calculation and evaluates the
ordering of those results.  And even though this operation may not
reveal the exact distribution of the mixing effects, it has important
practical implications, as it gives an indication of the scales of
those effects---not just their presence or absence.  This can be used,
together with expert knowledge and other data-analytic techniques, to
select an appropriate lower bound on the interval at which one should
analyze, or report, the data.

Departures from pure monotonicity in the $PE$ traces are a complex and
interesting question.  If $\overline{\R}$ does not reach zero for any bin
size, for instance, one can choose the first minimum of a curve like
the ones in Figure~\ref{fig:Rbar-vs-binsize} or~\ref{fig:rbar-binsize}
to perform the binning, but that is not guaranteed to remove all of
the mixing effects---especially if nonstationarities are involved,
causing the mixing scales to vary across the data set.  Deconvolving
multiple nonstationary mixing processes would call for strategies far
more sophisticated than simple binning at a fixed resolution across
the whole time series.  A formal solution to this would be a major
challenge, given the complex nature of the nonlinear statistics that
underpin permutation-entropy analyses.  Nonetheless, a windowed
$\overline{\R}$ analysis like the one shown in
Figure~\ref{fig:windowed-rbar} can help ascertain whether there are
different regimes in the data, with different noise \& mixing
properties in each---as in the different segments of the Mauna Loa
methane data, before and after the instrument change.

A major advantage of our technique is its model-free nature: it
identifies the mixing scale regardless of the cause.  This has
disadvantages as well, though: if two mixing processes are at work,
our analysis will bring out the larger one.  If the scales are widely
separated and one has very fine-grained data, it might be possible to
extend this analysis to detect the different scales.
This is a potential issue in the WDC data, where there are two known
sources of mixing: diffusion of the isotopes in the ice sheet and
fluid mixing in the laboratory analysis pipeline.  Diffusion is a
function of density, and density in a ice core is a function of depth,
so the effects are not simple.
Standard procedure in this laboratory is to use estimates of the
accumulation rate and temperature to make an educated guess at the
diffusion scale at each depth.  In constructing the published data set
from the raw data, the analysts chose a sample rate that exceeded that
resolution---in effect, intentionally oversampling the data---and then
selected a bin size that spaced the reported data points safely above
their estimates of the mixing scales.
If those estimates are good, this procedure will obviously make the
two mixing scales (diffusion in the ice sheet and fluid dynamics in
the laboratory apparatus) very similar, so we do not expect to be able
to deconvolve these two effects from this data set.  

Similarly, there are a number of different potential sources of the
mixing effects in the Mauna Loa data.
While the sample lines are unlikely to be at issue because the flow
through them can be approximated as plug flow, there is noise in the
analyzer.  There are also known sources of organized natural
variability in the study system itself, which span the time scales
from days to decades:
\begin{enumerate} 
\item Diurnal cycle related to local wind regime,
\item Synoptic scale variability occurring on time scales of weather
  fronts (a few days),
\item Seasonal cycle related to changing
atmospheric loss process rate, and 
\item Long-term trend related to
imbalance between emissions and losses.  
\end{enumerate}
We did not find any correlation between $\overline{\R}$ and meteorological
parameters, although Dlugokencky {\sl et al.} \cite{dlugokencky95}
showed that, on average, the diurnal cycle in atmospheric $CH_4$
correlated with dew point, because both are impacted by daily cycling
between upslope and downslope wind regimes.  Of course, there could be
convolved effects from multiple causes, but those would be difficult
to study in isolation with a permutation-entropy analysis for the same
reasons mentioned at the end of the previous paragraph.

The alignment between our results and scientific knowledge about the
experimental situation is encouraging.  The $\overline{\R}$ calculations
also clearly bring out the timing, as well as the salutary effects, of
the 2019 instrument change on Mauna Loa, for instance.  In the ice
core data, the $\overline{\R}$ metric revealed that a 15-point averaging
window restored the normal order of the $PE$ traces, when the experts
chose a 17-point window.  Similarly, our method suggested a one-hour
mixing window for the Mauna Loa data, which is exactly the cadence
that is used in practice.  In the data gathered by the newer
instrument, the same strategy indicated a significant reduction in the
mixing scales, corresponding to a 25-minute cadence.  

While corroboration of the known properties of a new instrument is not
that exciting, there can be unknown mechanisms at work in one's data,
leaving traces that are all but invisible to normal analysis
techniques.  Indeed, this was the driving force behind the creation of
the WDC data set used for this work.  After a permutation entropy
analysis revealed noise introduced by the CRDS instrument that had
been used to analyze this section of the ice core
\cite{garland2018anomaly}, the analysis was repeated using a newer
instrument, removing the noise effects.  Finally, the results of the
analysis described in the previous sections not only corroborate the
reasoning behind the approaches taken by the experts, but could also
provide a more systematic way to figure out exactly how to squeeze as
much information out of a study system as nature will allow.  Note
that this does not mean that data measured below the mixing scale are
useless; indeed, one must sample below a scale in order to identify
its lower boundary.

There are many avenues for extension of this work.  The metric used in
the pointwise $\tau$-reversal metric ($\R$) and the temporally
aggregated nature of its average, $\overline{\R}$, have pros and cons.
Recall that $\R$ is based solely on the L1 norm of the distance
between the focal $\tau$-sequence vector $\mathbf{v}$ and the purely
monotone vectors $\mathbf{v}_i$ and $\mathbf{v}_d$.  While this is a
sensible measure of an overall pattern of monotonicity, it does
incorporate some bias: an interchange of the $\tau=6$ and $\tau=1$
$PE$ traces, for instance, will affect the $\R$ value more strongly
than an interchange of the $\tau=3$ and $\tau=4$ traces.  And because
$\overline{\R}$ is an average of $\R$ across a segment of the data set, it
cannot capture temporal detail that occurs below the scale of the
segment, such as regimes of reversed and normal ordering at different
points in the data.  Since many real-world data sets are
nonstationary, it might be useful to develop a formalized strategy for
a windowed $\overline{\R}$ analysis with varying window sizes.  (The window
size in Figure~\ref{fig:windowed-rbar} tuned by hand to bring out the
operative effects without excessive detail.)  One could compute some
statistics on the lengths of reversed and normal regimes from these
results; even just the maxima and minima in $\overline{\R}$ might be
informative.  This could be particularly useful in ice cores, where
large diffusion events are theorized to occur \cite{JGRF:JGRF20648}.
In cases like this, these $\tau$-reversal metrics could provide
insight not only into the instruments and sampling rates that are
appropriate for different time scales, but also into the scientific
questions of a field.  Applications of the techniques described in
this paper are by no means limited to purely temporal data, either, as
evidenced by the WDC example.  Since mixing is an important issue in
spatial and spatio-temporal data sets, that is an additional advantage
of the technique, and an avenue for further investigation.

\section{Conclusion}\label{sec:concl}

In experimental practice, a delicate balance can arise between
properly sampling the dynamics of a study system and oversampling them
and thereby obfuscating the underlying signal with measurement-related
effects.  The Nyquist rate provides some guidance for this when the
system is linear and its highest frequency is known, but those
conditions are not always met in real-world situations.  In this
manuscript, we have described an information-theoretic technique for
detecting one such oversampling effect, which arises when local mixing
is present.
Critically, this method requires no domain-specific knowledge about
the data, or about the system that produced it.  One simply calculates
the permutation entropy across the time series at different temporal
resolutions and examines the relationship between the resulting
traces.

For most time series, permutation entropy will increase with the
stride of the calculation (the $\tau$ parameter in the previous
sections).  This simply reflects decreasing predictability over the
longer time span sampled by each permutation.  Reversal of that
ordering suggests that local mixing has occurred.  The basic intuition
here is straightforward: if successive data points are measured on a
smaller scale than the mixing scales that are inherent in the data,
each measurement is essentially a single draw from a local
distribution composed of the data points in its local neighborhood.
By definition, this added randomness will raise the entropy rate.  A
reversal of the normal ordering of the traces, then, is an indication
that mixing may be at work in the data.  One can extend this reasoning
to estimate the scale of the mixing effects via binned averaging,
which eliminates the effects of local mixing.  And mixing is not the
only data issue that $\overline{\R}$ can flag; we are currently
investigating rounding effects, for instance, which can also change
$PE$ values and reshuffle the $\tau$ ordering.

Our method can not only be used by practitioners to detect mixing in
their data; it can also help them choose appropriate lower bounds on
the scales at which to perform the analysis, allowing them to squeeze
as much information as possible out of their study system while
avoiding spurious effects.
Sampling frequency is often an imperfect balance between time, money,
scientific need, and what the system of interest allows.  Educated
guesses are frequently our only means of making decisions about how
often to sample.  Considerable time and money could be saved if we had
some meaningful feedback about where to draw the line in terms of
sampling.  Accuracy is also an issue here, since local mixing
obfuscates data; in a situation like this, knowing the scales at which
one should perform an analysis is of obvious value.  Moreover, a
method that identifies oversampling scales can be used to establish
solid guidelines for data reporting.  And our method does not only
work on {\sl time} series data, as is clear from our ice-core example,
where the $\tau$-reversal metric revealed mixing on spatial scales in
the core that aligned with expert reasoning about diffusion in the ice
sheet.

Permutation entropy and its variants (e.g.,\cite{fadlallah2013}) have
become a staple in time-series analysis, especially in anomaly
detection and forecasting.  Some of that work has leveraged the $\tau$
parameter to focus on different temporal scales.  To the best of our
knowledge, though, no one has used the ordering of $PE$ traces
calculated across different $\tau$ values to understand the properties
of the data.

\clearpage

\bibliography{refs}

\begin{thebibliography}{34}%
\makeatletter
\providecommand \@ifxundefined [1]{%
 \@ifx{#1\undefined}
}%
\providecommand \@ifnum [1]{%
 \ifnum #1\expandafter \@firstoftwo
 \else \expandafter \@secondoftwo
 \fi
}%
\providecommand \@ifx [1]{%
 \ifx #1\expandafter \@firstoftwo
 \else \expandafter \@secondoftwo
 \fi
}%
\providecommand \natexlab [1]{#1}%
\providecommand \enquote  [1]{``#1''}%
\providecommand \bibnamefont  [1]{#1}%
\providecommand \bibfnamefont [1]{#1}%
\providecommand \citenamefont [1]{#1}%
\providecommand \href@noop [0]{\@secondoftwo}%
\providecommand \href [0]{\begingroup \@sanitize@url \@href}%
\providecommand \@href[1]{\@@startlink{#1}\@@href}%
\providecommand \@@href[1]{\endgroup#1\@@endlink}%
\providecommand \@sanitize@url [0]{\catcode `\\12\catcode `\$12\catcode
  `\&12\catcode `\#12\catcode `\^12\catcode `\_12\catcode `\%12\relax}%
\providecommand \@@startlink[1]{}%
\providecommand \@@endlink[0]{}%
\providecommand \url  [0]{\begingroup\@sanitize@url \@url }%
\providecommand \@url [1]{\endgroup\@href {#1}{\urlprefix }}%
\providecommand \urlprefix  [0]{URL }%
\providecommand \Eprint [0]{\href }%
\providecommand \doibase [0]{https://doi.org/}%
\providecommand \selectlanguage [0]{\@gobble}%
\providecommand \bibinfo  [0]{\@secondoftwo}%
\providecommand \bibfield  [0]{\@secondoftwo}%
\providecommand \translation [1]{[#1]}%
\providecommand \BibitemOpen [0]{}%
\providecommand \bibitemStop [0]{}%
\providecommand \bibitemNoStop [0]{.\EOS\space}%
\providecommand \EOS [0]{\spacefactor3000\relax}%
\providecommand \BibitemShut  [1]{\csname bibitem#1\endcsname}%
\let\auto@bib@innerbib\@empty
%</preamble>
\bibitem [{\citenamefont {Bandt}\ and\ \citenamefont
  {Pompe}(2002)}]{bandt2002permutation}%
  \BibitemOpen
  \bibfield  {author} {\bibinfo {author} {\bibfnamefont {C.}~\bibnamefont
  {Bandt}}\ and\ \bibinfo {author} {\bibfnamefont {B.}~\bibnamefont {Pompe}},\
  }\bibfield  {title} {\bibinfo {title} {Permutation entropy: {A} natural
  complexity measure for time series},\ }\href@noop {} {\bibfield  {journal}
  {\bibinfo  {journal} {Physical Review Letters}\ }\textbf {\bibinfo {volume}
  {88}},\ \bibinfo {pages} {174102} (\bibinfo {year} {2002})}\BibitemShut
  {NoStop}%
\bibitem [{\citenamefont {Amig{\'o}}(2010)}]{amigo2010permutation}%
  \BibitemOpen
  \bibfield  {author} {\bibinfo {author} {\bibfnamefont {J.}~\bibnamefont
  {Amig{\'o}}},\ }\href@noop {} {\emph {\bibinfo {title} {Permutation
  Complexity in Dynamical Systems: Ordinal Patterns, Permutation Entropy and
  All That}}}\ (\bibinfo  {publisher} {Springer Science \& Business Media},\
  \bibinfo {year} {2010})\BibitemShut {NoStop}%
\bibitem [{\citenamefont {Pennekamp}\ \emph {et~al.}(2019)\citenamefont
  {Pennekamp}, \citenamefont {Iles}, \citenamefont {Garland}, \citenamefont
  {Brennan}, \citenamefont {Brose}, \citenamefont {Gaedke}, \citenamefont
  {Jacob}, \citenamefont {Kratina}, \citenamefont {Matthews}, \citenamefont
  {Munch} \emph {et~al.}}]{pennekamp2019intrinsic}%
  \BibitemOpen
  \bibfield  {author} {\bibinfo {author} {\bibfnamefont {F.}~\bibnamefont
  {Pennekamp}}, \bibinfo {author} {\bibfnamefont {A.~C.}\ \bibnamefont {Iles}},
  \bibinfo {author} {\bibfnamefont {J.}~\bibnamefont {Garland}}, \bibinfo
  {author} {\bibfnamefont {G.}~\bibnamefont {Brennan}}, \bibinfo {author}
  {\bibfnamefont {U.}~\bibnamefont {Brose}}, \bibinfo {author} {\bibfnamefont
  {U.}~\bibnamefont {Gaedke}}, \bibinfo {author} {\bibfnamefont
  {U.}~\bibnamefont {Jacob}}, \bibinfo {author} {\bibfnamefont
  {P.}~\bibnamefont {Kratina}}, \bibinfo {author} {\bibfnamefont
  {B.}~\bibnamefont {Matthews}}, \bibinfo {author} {\bibfnamefont
  {S.}~\bibnamefont {Munch}}, \emph {et~al.},\ }\bibfield  {title} {\bibinfo
  {title} {The intrinsic predictability of ecological time series and its
  potential to guide forecasting},\ }\href@noop {} {\bibfield  {journal}
  {\bibinfo  {journal} {Ecological Monographs}\ }\textbf {\bibinfo {volume}
  {89}},\ \bibinfo {pages} {e01359} (\bibinfo {year} {2019})}\BibitemShut
  {NoStop}%
\bibitem [{\citenamefont {Garland}\ \emph {et~al.}(2014)\citenamefont
  {Garland}, \citenamefont {James},\ and\ \citenamefont
  {Bradley}}]{garland2014model}%
  \BibitemOpen
  \bibfield  {author} {\bibinfo {author} {\bibfnamefont {J.}~\bibnamefont
  {Garland}}, \bibinfo {author} {\bibfnamefont {R.}~\bibnamefont {James}},\
  and\ \bibinfo {author} {\bibfnamefont {E.}~\bibnamefont {Bradley}},\
  }\bibfield  {title} {\bibinfo {title} {Model-free quantification of
  time-series predictability},\ }\href@noop {} {\bibfield  {journal} {\bibinfo
  {journal} {Physical Review E}\ }\textbf {\bibinfo {volume} {90}},\ \bibinfo
  {pages} {052910} (\bibinfo {year} {2014})}\BibitemShut {NoStop}%
\bibitem [{\citenamefont {Scarpino}\ and\ \citenamefont
  {Petri}(2019)}]{scarpino2019predictability}%
  \BibitemOpen
  \bibfield  {author} {\bibinfo {author} {\bibfnamefont {S.~V.}\ \bibnamefont
  {Scarpino}}\ and\ \bibinfo {author} {\bibfnamefont {G.}~\bibnamefont
  {Petri}},\ }\bibfield  {title} {\bibinfo {title} {On the predictability of
  infectious disease outbreaks},\ }\href@noop {} {\bibfield  {journal}
  {\bibinfo  {journal} {Nature Communications}\ }\textbf {\bibinfo {volume}
  {10}},\ \bibinfo {pages} {1} (\bibinfo {year} {2019})}\BibitemShut {NoStop}%
\bibitem [{Note1()}]{Note1}%
  \BibitemOpen
  \bibinfo {note} {The integration step size used to generate the time series
  also plays a role here; if it is very small, successive points are highly
  correlated and the $PE$ will be lower.}\BibitemShut {Stop}%
\bibitem [{\citenamefont {Myers}\ and\ \citenamefont
  {Khasawneh}(2020)}]{myers2020automatic}%
  \BibitemOpen
  \bibfield  {author} {\bibinfo {author} {\bibfnamefont {A.}~\bibnamefont
  {Myers}}\ and\ \bibinfo {author} {\bibfnamefont {F.~A.}\ \bibnamefont
  {Khasawneh}},\ }\bibfield  {title} {\bibinfo {title} {On the automatic
  parameter selection for permutation entropy},\ }\href@noop {} {\bibfield
  {journal} {\bibinfo  {journal} {Chaos}\ }\textbf {\bibinfo {volume} {30}},\
  \bibinfo {pages} {033130} (\bibinfo {year} {2020})}\BibitemShut {NoStop}%
\bibitem [{\citenamefont {Riedl}\ \emph {et~al.}(2013)\citenamefont {Riedl},
  \citenamefont {M{\"u}ller},\ and\ \citenamefont
  {Wessel}}]{riedl2013practical}%
  \BibitemOpen
  \bibfield  {author} {\bibinfo {author} {\bibfnamefont {M.}~\bibnamefont
  {Riedl}}, \bibinfo {author} {\bibfnamefont {A.}~\bibnamefont {M{\"u}ller}},\
  and\ \bibinfo {author} {\bibfnamefont {N.}~\bibnamefont {Wessel}},\
  }\bibfield  {title} {\bibinfo {title} {Practical considerations of
  permutation entropy},\ }\href@noop {} {\bibfield  {journal} {\bibinfo
  {journal} {The European Physical Journal Special Topics}\ }\textbf {\bibinfo
  {volume} {222}},\ \bibinfo {pages} {249} (\bibinfo {year}
  {2013})}\BibitemShut {NoStop}%
\bibitem [{\citenamefont {Cao}\ \emph {et~al.}(2004)\citenamefont {Cao},
  \citenamefont {Tung}, \citenamefont {Gao}, \citenamefont {Protopopescu},\
  and\ \citenamefont {Hively}}]{cao2004detecting}%
  \BibitemOpen
  \bibfield  {author} {\bibinfo {author} {\bibfnamefont {Y.}~\bibnamefont
  {Cao}}, \bibinfo {author} {\bibfnamefont {W.-w.}\ \bibnamefont {Tung}},
  \bibinfo {author} {\bibfnamefont {J.}~\bibnamefont {Gao}}, \bibinfo {author}
  {\bibfnamefont {V.~A.}\ \bibnamefont {Protopopescu}},\ and\ \bibinfo {author}
  {\bibfnamefont {L.~M.}\ \bibnamefont {Hively}},\ }\bibfield  {title}
  {\bibinfo {title} {Detecting dynamical changes in time series using the
  permutation entropy},\ }\href@noop {} {\bibfield  {journal} {\bibinfo
  {journal} {Physical Review E}\ }\textbf {\bibinfo {volume} {70}},\ \bibinfo
  {pages} {046217} (\bibinfo {year} {2004})}\BibitemShut {NoStop}%
\bibitem [{\citenamefont {Garland}\ \emph {et~al.}(2019)\citenamefont
  {Garland}, \citenamefont {Jones}, \citenamefont {Neuder}, \citenamefont
  {White},\ and\ \citenamefont {Bradley}}]{garland2019information}%
  \BibitemOpen
  \bibfield  {author} {\bibinfo {author} {\bibfnamefont {J.}~\bibnamefont
  {Garland}}, \bibinfo {author} {\bibfnamefont {T.~R.}\ \bibnamefont {Jones}},
  \bibinfo {author} {\bibfnamefont {M.}~\bibnamefont {Neuder}}, \bibinfo
  {author} {\bibfnamefont {J.~W.}\ \bibnamefont {White}},\ and\ \bibinfo
  {author} {\bibfnamefont {E.}~\bibnamefont {Bradley}},\ }\bibfield  {title}
  {\bibinfo {title} {An information-theoretic approach to extracting climate
  signals from deep polar ice cores},\ }\href@noop {} {\bibfield  {journal}
  {\bibinfo  {journal} {Chaos}\ }\textbf {\bibinfo {volume} {29}},\ \bibinfo
  {pages} {101105} (\bibinfo {year} {2019})}\BibitemShut {NoStop}%
\bibitem [{\citenamefont {Garland}\ \emph {et~al.}(2018)\citenamefont
  {Garland}, \citenamefont {Jones}, \citenamefont {Neuder}, \citenamefont
  {Morris}, \citenamefont {White},\ and\ \citenamefont
  {Bradley}}]{garland2018anomaly}%
  \BibitemOpen
  \bibfield  {author} {\bibinfo {author} {\bibfnamefont {J.}~\bibnamefont
  {Garland}}, \bibinfo {author} {\bibfnamefont {T.~R.}\ \bibnamefont {Jones}},
  \bibinfo {author} {\bibfnamefont {M.}~\bibnamefont {Neuder}}, \bibinfo
  {author} {\bibfnamefont {V.}~\bibnamefont {Morris}}, \bibinfo {author}
  {\bibfnamefont {J.~W.}\ \bibnamefont {White}},\ and\ \bibinfo {author}
  {\bibfnamefont {E.}~\bibnamefont {Bradley}},\ }\bibfield  {title} {\bibinfo
  {title} {Anomaly detection in paleoclimate records using permutation
  entropy},\ }\href@noop {} {\bibfield  {journal} {\bibinfo  {journal}
  {Entropy}\ }\textbf {\bibinfo {volume} {20}},\ \bibinfo {pages} {931}
  (\bibinfo {year} {2018})}\BibitemShut {NoStop}%
\bibitem [{\citenamefont {Amig{\'{o}}}\ \emph {et~al.}(2008)\citenamefont
  {Amig{\'{o}}}, \citenamefont {Zambrano},\ and\ \citenamefont
  {Sanju{\'{a}}n}}]{Amigo2008}%
  \BibitemOpen
  \bibfield  {author} {\bibinfo {author} {\bibfnamefont {J.~M.}\ \bibnamefont
  {Amig{\'{o}}}}, \bibinfo {author} {\bibfnamefont {S.}~\bibnamefont
  {Zambrano}},\ and\ \bibinfo {author} {\bibfnamefont {M.~A.~F.}\ \bibnamefont
  {Sanju{\'{a}}n}},\ }\bibfield  {title} {\bibinfo {title} {Combinatorial
  detection of determinism in noisy time series},\ }\href
  {https://doi.org/10.1209/0295-5075/83/60005} {\bibfield  {journal} {\bibinfo
  {journal} {Europhysics Letters}\ }\textbf {\bibinfo {volume} {83}},\ \bibinfo
  {pages} {60005} (\bibinfo {year} {2008})}\BibitemShut {NoStop}%
\bibitem [{\citenamefont {Garland}\ \emph {et~al.}(2016)\citenamefont
  {Garland}, \citenamefont {Jones}, \citenamefont {Bradley}, \citenamefont
  {James},\ and\ \citenamefont {White}}]{garland2016first}%
  \BibitemOpen
  \bibfield  {author} {\bibinfo {author} {\bibfnamefont {J.}~\bibnamefont
  {Garland}}, \bibinfo {author} {\bibfnamefont {T.~R.}\ \bibnamefont {Jones}},
  \bibinfo {author} {\bibfnamefont {E.}~\bibnamefont {Bradley}}, \bibinfo
  {author} {\bibfnamefont {R.~G.}\ \bibnamefont {James}},\ and\ \bibinfo
  {author} {\bibfnamefont {J.~W.}\ \bibnamefont {White}},\ }\bibfield  {title}
  {\bibinfo {title} {A first step toward quantifying the climate’s
  information production over the last 68,000 years},\ }in\ \href@noop {}
  {\emph {\bibinfo {booktitle} {International Symposium on Intelligent Data
  Analysis}}}\ (\bibinfo {organization} {Springer},\ \bibinfo {year} {2016})\
  pp.\ \bibinfo {pages} {343--355}\BibitemShut {NoStop}%
\bibitem [{\citenamefont {Myktowicz}\ \emph {et~al.}(2009)\citenamefont
  {Myktowicz}, \citenamefont {Diwan},\ and\ \citenamefont
  {Bradley}}]{myktowicz2009}%
  \BibitemOpen
  \bibfield  {author} {\bibinfo {author} {\bibfnamefont {T.}~\bibnamefont
  {Myktowicz}}, \bibinfo {author} {\bibfnamefont {A.}~\bibnamefont {Diwan}},\
  and\ \bibinfo {author} {\bibfnamefont {E.}~\bibnamefont {Bradley}},\
  }\bibfield  {title} {\bibinfo {title} {Computers are dynamical systems},\
  }\href@noop {} {\bibfield  {journal} {\bibinfo  {journal} {Chaos}\ }\textbf
  {\bibinfo {volume} {19}},\ \bibinfo {pages} {033124} (\bibinfo {year}
  {2009})}\BibitemShut {NoStop}%
\bibitem [{\citenamefont {Garland}\ and\ \citenamefont
  {Bradley}(2011)}]{garland2011predicting}%
  \BibitemOpen
  \bibfield  {author} {\bibinfo {author} {\bibfnamefont {J.}~\bibnamefont
  {Garland}}\ and\ \bibinfo {author} {\bibfnamefont {E.}~\bibnamefont
  {Bradley}},\ }\bibfield  {title} {\bibinfo {title} {Predicting computer
  performance dynamics},\ }in\ \href@noop {} {\emph {\bibinfo {booktitle}
  {International Symposium on Intelligent Data Analysis}}}\ (\bibinfo
  {organization} {Springer},\ \bibinfo {year} {2011})\ pp.\ \bibinfo {pages}
  {173--184}\BibitemShut {NoStop}%
\bibitem [{\citenamefont {Bradley}()}]{liz-pend-data}%
  \BibitemOpen
  \bibfield  {author} {\bibinfo {author} {\bibfnamefont {E.}~\bibnamefont
  {Bradley}},\ }\href {https://doi.org/https://bit.ly/2UAuADp} {\bibinfo
  {title} {Damped-driven pendulum dataset}},\ \bibinfo {howpublished}
  {https://bit.ly/2UAuADp}\BibitemShut {NoStop}%
\bibitem [{\citenamefont {Diaconis}\ and\ \citenamefont
  {Graham}(1977)}]{diaconis1977spearman}%
  \BibitemOpen
  \bibfield  {author} {\bibinfo {author} {\bibfnamefont {P.}~\bibnamefont
  {Diaconis}}\ and\ \bibinfo {author} {\bibfnamefont {R.~L.}\ \bibnamefont
  {Graham}},\ }\bibfield  {title} {\bibinfo {title} {Spearman's footrule as a
  measure of disarray},\ }\href@noop {} {\bibfield  {journal} {\bibinfo
  {journal} {Journal of the Royal Statistical Society: Series B
  (Methodological)}\ }\textbf {\bibinfo {volume} {39}},\ \bibinfo {pages} {262}
  (\bibinfo {year} {1977})}\BibitemShut {NoStop}%
\bibitem [{\citenamefont {Kendall}(1948)}]{kendall1948rank}%
  \BibitemOpen
  \bibfield  {author} {\bibinfo {author} {\bibfnamefont {M.~G.}\ \bibnamefont
  {Kendall}},\ }\bibfield  {title} {\bibinfo {title} {Rank correlation
  methods.},\ }\href@noop {} {\  (\bibinfo {year} {1948})}\BibitemShut
  {NoStop}%
\bibitem [{\citenamefont {Abdi}(2007)}]{abdi2007kendall}%
  \BibitemOpen
  \bibfield  {author} {\bibinfo {author} {\bibfnamefont {H.}~\bibnamefont
  {Abdi}},\ }\bibfield  {title} {\bibinfo {title} {The kendall rank correlation
  coefficient},\ }\href@noop {} {\bibfield  {journal} {\bibinfo  {journal}
  {Encyclopedia of Measurement and Statistics. Sage, Thousand Oaks, CA}\ ,\
  \bibinfo {pages} {508}} (\bibinfo {year} {2007})}\BibitemShut {NoStop}%
\bibitem [{\citenamefont {Lorenz}(1963)}]{lorenz1963deterministic}%
  \BibitemOpen
  \bibfield  {author} {\bibinfo {author} {\bibfnamefont {E.~N.}\ \bibnamefont
  {Lorenz}},\ }\bibfield  {title} {\bibinfo {title} {Deterministic nonperiodic
  flow},\ }\href@noop {} {\bibfield  {journal} {\bibinfo  {journal} {Journal of
  the Atmospheric Sciences}\ }\textbf {\bibinfo {volume} {20}},\ \bibinfo
  {pages} {130} (\bibinfo {year} {1963})}\BibitemShut {NoStop}%
\bibitem [{Note2()}]{Note2}%
  \BibitemOpen
  \bibinfo {note} {For systems with delays, such as those modeled with
  delay-differential equations, a wider binning window may be required because
  of the associated temporal propagation of information due to the delay
  term.}\BibitemShut {Stop}%
\bibitem [{\citenamefont {Mackey}\ and\ \citenamefont
  {Glass}(1977)}]{Mackey287}%
  \BibitemOpen
  \bibfield  {author} {\bibinfo {author} {\bibfnamefont {M.}~\bibnamefont
  {Mackey}}\ and\ \bibinfo {author} {\bibfnamefont {L.}~\bibnamefont {Glass}},\
  }\bibfield  {title} {\bibinfo {title} {Oscillation and chaos in physiological
  control systems},\ }\href {https://doi.org/10.1126/science.267326} {\bibfield
   {journal} {\bibinfo  {journal} {Science}\ }\textbf {\bibinfo {volume}
  {197}},\ \bibinfo {pages} {287} (\bibinfo {year} {1977})},\ \Eprint
  {https://arxiv.org/abs/https://science.sciencemag.org/content/197/4300/287.full.pdf}
  {https://science.sciencemag.org/content/197/4300/287.full.pdf} \BibitemShut
  {NoStop}%
\bibitem [{uci()}]{uciMLrepo}%
  \BibitemOpen
  \href@noop {} {\bibinfo {title} {{UCI} machine learning repository}},\
  \bibinfo {note} {{\tt http://archive.ics.uci.edu/ml}}\BibitemShut {NoStop}%
\bibitem [{\citenamefont {Fonollosa}\ \emph {et~al.}(2015)\citenamefont
  {Fonollosa}, \citenamefont {Sheik}, \citenamefont {Huerta},\ and\
  \citenamefont {Marco}}]{fonollosa2015reservoir}%
  \BibitemOpen
  \bibfield  {author} {\bibinfo {author} {\bibfnamefont {J.}~\bibnamefont
  {Fonollosa}}, \bibinfo {author} {\bibfnamefont {S.}~\bibnamefont {Sheik}},
  \bibinfo {author} {\bibfnamefont {R.}~\bibnamefont {Huerta}},\ and\ \bibinfo
  {author} {\bibfnamefont {S.}~\bibnamefont {Marco}},\ }\bibfield  {title}
  {\bibinfo {title} {Reservoir computing compensates slow response of
  chemosensor arrays exposed to fast varying gas concentrations in continuous
  monitoring},\ }\href@noop {} {\bibfield  {journal} {\bibinfo  {journal}
  {Sensors and Actuators B: Chemical}\ }\textbf {\bibinfo {volume} {215}},\
  \bibinfo {pages} {618} (\bibinfo {year} {2015})}\BibitemShut {NoStop}%
\bibitem [{Note3()}]{Note3}%
  \BibitemOpen
  \bibinfo {note} {We do not have enough data to extend the calculation beyond
  $j=60$.}\BibitemShut {Stop}%
\bibitem [{\citenamefont {Dansgaard}(1964)}]{Dansgaard-64-Tellus}%
  \BibitemOpen
  \bibfield  {author} {\bibinfo {author} {\bibfnamefont {W.}~\bibnamefont
  {Dansgaard}},\ }\bibfield  {title} {\bibinfo {title} {Stable isotopes in
  precipitation},\ }\href {https://doi.org/10.1111/j.2153-3490.1964.tb00181.x}
  {\bibfield  {journal} {\bibinfo  {journal} {Tellus}\ }\textbf {\bibinfo
  {volume} {16}},\ \bibinfo {pages} {436} (\bibinfo {year} {1964})},\ \Eprint
  {https://arxiv.org/abs/https://onlinelibrary.wiley.com/doi/pdf/10.1111/j.2153-3490.1964.tb00181.x}
  {https://onlinelibrary.wiley.com/doi/pdf/10.1111/j.2153-3490.1964.tb00181.x}
  \BibitemShut {NoStop}%
\bibitem [{\citenamefont {Jones}\ \emph
  {et~al.}(2017{\natexlab{a}})\citenamefont {Jones}, \citenamefont {White},
  \citenamefont {Steig}, \citenamefont {Vaughn}, \citenamefont {Morris},
  \citenamefont {Gkinis}, \citenamefont {Markle},\ and\ \citenamefont
  {Schoenemann}}]{jones2017improved}%
  \BibitemOpen
  \bibfield  {author} {\bibinfo {author} {\bibfnamefont {T.~R.}\ \bibnamefont
  {Jones}}, \bibinfo {author} {\bibfnamefont {J.~W.}\ \bibnamefont {White}},
  \bibinfo {author} {\bibfnamefont {E.~J.}\ \bibnamefont {Steig}}, \bibinfo
  {author} {\bibfnamefont {B.~H.}\ \bibnamefont {Vaughn}}, \bibinfo {author}
  {\bibfnamefont {V.}~\bibnamefont {Morris}}, \bibinfo {author} {\bibfnamefont
  {V.}~\bibnamefont {Gkinis}}, \bibinfo {author} {\bibfnamefont {B.~R.}\
  \bibnamefont {Markle}},\ and\ \bibinfo {author} {\bibfnamefont {S.~W.}\
  \bibnamefont {Schoenemann}},\ }\bibfield  {title} {\bibinfo {title} {Improved
  methodologies for continuous-flow analysis of stable water isotopes in ice
  cores},\ }\href@noop {} {\bibfield  {journal} {\bibinfo  {journal}
  {Atmospheric Measurement Techniques}\ }\textbf {\bibinfo {volume} {10}}
  (\bibinfo {year} {2017}{\natexlab{a}})}\BibitemShut {NoStop}%
\bibitem [{\citenamefont {Jones}\ \emph {et~al.}()\citenamefont {Jones},
  \citenamefont {Garland}, \citenamefont {Morris}, \citenamefont {Neuder},
  \citenamefont {Bradley},\ and\ \citenamefont {White}}]{WDCReSample}%
  \BibitemOpen
  \bibfield  {author} {\bibinfo {author} {\bibfnamefont {T.~R.}\ \bibnamefont
  {Jones}}, \bibinfo {author} {\bibfnamefont {J.}~\bibnamefont {Garland}},
  \bibinfo {author} {\bibfnamefont {V.}~\bibnamefont {Morris}}, \bibinfo
  {author} {\bibfnamefont {M.}~\bibnamefont {Neuder}}, \bibinfo {author}
  {\bibfnamefont {E.}~\bibnamefont {Bradley}},\ and\ \bibinfo {author}
  {\bibfnamefont {J.~W.~C.}\ \bibnamefont {White}},\ }\href
  {https://doi.org/10.15784/601274} {\bibinfo {title} {Resampling of deep polar
  ice cores using information theory}},\ \bibinfo {howpublished} {U.S.
  Antarctic Program (USAP) Data Center. doi: 10.15784/601274}\BibitemShut
  {NoStop}%
\bibitem [{\citenamefont {Dlugokencky}\ \emph {et~al.}(1995)\citenamefont
  {Dlugokencky}, \citenamefont {Steele}, \citenamefont {Lang},\ and\
  \citenamefont {Masarie}}]{dlugokencky95}%
  \BibitemOpen
  \bibfield  {author} {\bibinfo {author} {\bibfnamefont {E.~J.}\ \bibnamefont
  {Dlugokencky}}, \bibinfo {author} {\bibfnamefont {L.~P.}\ \bibnamefont
  {Steele}}, \bibinfo {author} {\bibfnamefont {P.~M.}\ \bibnamefont {Lang}},\
  and\ \bibinfo {author} {\bibfnamefont {K.~A.}\ \bibnamefont {Masarie}},\
  }\bibfield  {title} {\bibinfo {title} {Atmospheric methane at mauna loa and
  barrow observatories: Presentation and analysis of in situ measurements},\
  }\href {https://doi.org/10.1029/95JD02460} {\bibfield  {journal} {\bibinfo
  {journal} {Journal of Geophysical Research: Atmospheres}\ }\textbf {\bibinfo
  {volume} {100}},\ \bibinfo {pages} {23103} (\bibinfo {year} {1995})},\
  \Eprint
  {https://arxiv.org/abs/https://agupubs.onlinelibrary.wiley.com/doi/pdf/10.1029/95JD02460}
  {https://agupubs.onlinelibrary.wiley.com/doi/pdf/10.1029/95JD02460}
  \BibitemShut {NoStop}%
\bibitem [{Note4()}]{Note4}%
  \BibitemOpen
  \bibinfo {note} {$10^-9$ mol $CH_4$ per mol dry air.}\BibitemShut {Stop}%
\bibitem [{\citenamefont {McCullough}\ \emph {et~al.}(2016)\citenamefont
  {McCullough}, \citenamefont {Sakellariou}, \citenamefont {Stemler},\ and\
  \citenamefont {Small}}]{mccullough2016counting}%
  \BibitemOpen
  \bibfield  {author} {\bibinfo {author} {\bibfnamefont {M.}~\bibnamefont
  {McCullough}}, \bibinfo {author} {\bibfnamefont {K.}~\bibnamefont
  {Sakellariou}}, \bibinfo {author} {\bibfnamefont {T.}~\bibnamefont
  {Stemler}},\ and\ \bibinfo {author} {\bibfnamefont {M.}~\bibnamefont
  {Small}},\ }\bibfield  {title} {\bibinfo {title} {Counting forbidden patterns
  in irregularly sampled time series. i. the effects of under-sampling, random
  depletion, and timing jitter},\ }\href@noop {} {\bibfield  {journal}
  {\bibinfo  {journal} {Chaos: An Interdisciplinary Journal of Nonlinear
  Science}\ }\textbf {\bibinfo {volume} {26}},\ \bibinfo {pages} {123103}
  (\bibinfo {year} {2016})}\BibitemShut {NoStop}%
\bibitem [{\citenamefont {Sakellariou}\ \emph {et~al.}(2016)\citenamefont
  {Sakellariou}, \citenamefont {McCullough}, \citenamefont {Stemler},\ and\
  \citenamefont {Small}}]{sakellariou2016counting}%
  \BibitemOpen
  \bibfield  {author} {\bibinfo {author} {\bibfnamefont {K.}~\bibnamefont
  {Sakellariou}}, \bibinfo {author} {\bibfnamefont {M.}~\bibnamefont
  {McCullough}}, \bibinfo {author} {\bibfnamefont {T.}~\bibnamefont
  {Stemler}},\ and\ \bibinfo {author} {\bibfnamefont {M.}~\bibnamefont
  {Small}},\ }\bibfield  {title} {\bibinfo {title} {Counting forbidden patterns
  in irregularly sampled time series. ii. reliability in the presence of highly
  irregular sampling},\ }\href@noop {} {\bibfield  {journal} {\bibinfo
  {journal} {Chaos: An Interdisciplinary Journal of Nonlinear Science}\
  }\textbf {\bibinfo {volume} {26}},\ \bibinfo {pages} {123104} (\bibinfo
  {year} {2016})}\BibitemShut {NoStop}%
\bibitem [{\citenamefont {Jones}\ \emph
  {et~al.}(2017{\natexlab{b}})\citenamefont {Jones}, \citenamefont {Cuffey},
  \citenamefont {White}, \citenamefont {Steig}, \citenamefont {Buizert},
  \citenamefont {Markle}, \citenamefont {McConnell},\ and\ \citenamefont
  {Sigl}}]{JGRF:JGRF20648}%
  \BibitemOpen
  \bibfield  {author} {\bibinfo {author} {\bibfnamefont {T.~R.}\ \bibnamefont
  {Jones}}, \bibinfo {author} {\bibfnamefont {K.~M.}\ \bibnamefont {Cuffey}},
  \bibinfo {author} {\bibfnamefont {J.~W.~C.}\ \bibnamefont {White}}, \bibinfo
  {author} {\bibfnamefont {E.~J.}\ \bibnamefont {Steig}}, \bibinfo {author}
  {\bibfnamefont {C.}~\bibnamefont {Buizert}}, \bibinfo {author} {\bibfnamefont
  {B.~R.}\ \bibnamefont {Markle}}, \bibinfo {author} {\bibfnamefont {J.~R.}\
  \bibnamefont {McConnell}},\ and\ \bibinfo {author} {\bibfnamefont
  {M.}~\bibnamefont {Sigl}},\ }\bibfield  {title} {\bibinfo {title} {Water
  isotope diffusion in the {WAIS} {Divide} ice core during the {Holocene} and
  last glacial},\ }\href {https://doi.org/10.1002/2016JF003938} {\bibfield
  {journal} {\bibinfo  {journal} {Journal of Geophysical Research: Earth
  Surface}\ }\textbf {\bibinfo {volume} {122}},\ \bibinfo {pages} {290}
  (\bibinfo {year} {2017}{\natexlab{b}})}\BibitemShut {NoStop}%
\bibitem [{\citenamefont {Fadlallah}\ \emph {et~al.}(2013)\citenamefont
  {Fadlallah}, \citenamefont {Chen}, \citenamefont {Keil},\ and\ \citenamefont
  {Pr{\'\i}ncipe}}]{fadlallah2013}%
  \BibitemOpen
  \bibfield  {author} {\bibinfo {author} {\bibfnamefont {B.}~\bibnamefont
  {Fadlallah}}, \bibinfo {author} {\bibfnamefont {B.}~\bibnamefont {Chen}},
  \bibinfo {author} {\bibfnamefont {A.}~\bibnamefont {Keil}},\ and\ \bibinfo
  {author} {\bibfnamefont {J.}~\bibnamefont {Pr{\'\i}ncipe}},\ }\bibfield
  {title} {\bibinfo {title} {Weighted-permutation entropy: A complexity measure
  for time series incorporating amplitude information},\ }\href@noop {}
  {\bibfield  {journal} {\bibinfo  {journal} {Physical Review E}\ }\textbf
  {\bibinfo {volume} {87}},\ \bibinfo {pages} {022911} (\bibinfo {year}
  {2013})}\BibitemShut {NoStop}%
\end{thebibliography}%

\clearpage

\subsection*{Acknowledgments}
\noindent \textbf{Author Contributions} MN, EB, and JG crafted the
research questions.  MN, EB, and JG designed the analyses. MN and JG
conducted the analyses. MN, EB, JW, ED, and JG wrote the manuscript and
contributed to discussion and idea development.\\
\noindent \textbf{Funding:} This research was funded by US National
Science Foundation (NSF) grant number 1807478.  JG was also partially
supported by an Applied Complexity Fellowship at the Santa Fe
Institute.\\
\noindent \textbf{Competing Interests} The authors declare that they
have no competing financial interests.\\
\noindent \textbf{Data and materials availability:} The gas mixture
data is located in the UCI Machine Learning repository,
\href{https://archive.ics.uci.edu/ml/machine-learning-databases/00322/}{archive.ics.uci.edu/ml/machine-learning-databases/00322/}.
The ice core data is located in the U.S. Antarctic Program Data
Center,
\href{https://doi.org/10.15784/601274}{doi.org/10.15784/601274}.  The
hourly averages of the Mauna Loa data can be found at
\begin{verbatim}
 ftp://aftp.cmdl.noaa.gov/data/trace_gases/ch4/in-situ/surface/
\end{verbatim}

\newpage
\beginsupplement

\section*{Supplementary Figures}
\begin{figure}[h]
    \centering
\begin{subfigure}{0.49\textwidth}
    \centering 
\includegraphics[width=1\textwidth]{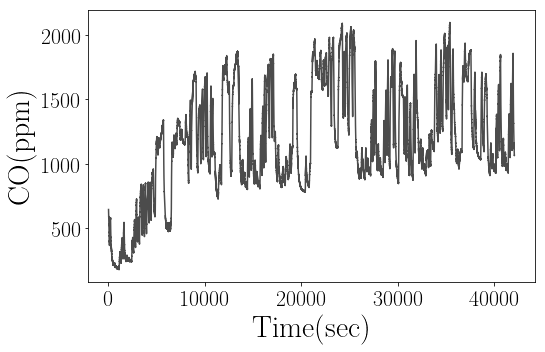}
\caption{Gas mixture concentrations -- Raw}
\end{subfigure}
\begin{subfigure}{0.49\textwidth}
    \centering 
\includegraphics[width=1\textwidth]{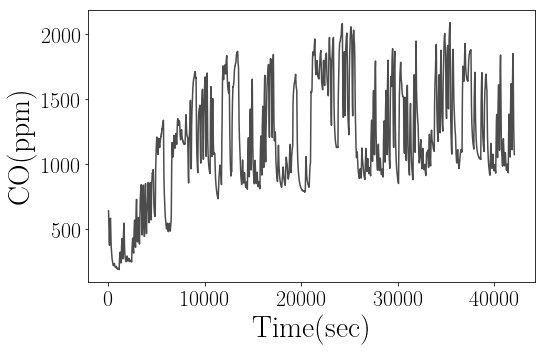}
\caption{Gas mixture concentrations -- Binned with $j=15$}
\end{subfigure}

\begin{subfigure}{0.49\textwidth}
    \centering 
\includegraphics[width=1\textwidth]{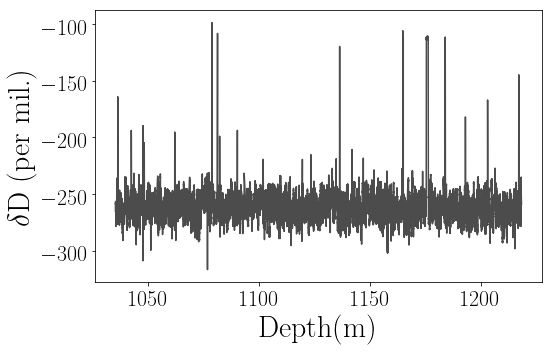}
\caption{WDC $\delta \text{D}$ -- Raw}
\end{subfigure}
\begin{subfigure}{0.49\textwidth}
    \centering 
\includegraphics[width=1\textwidth]{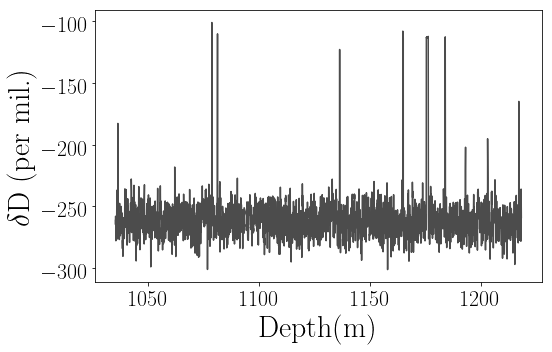}
\caption{WDC $\delta \text{D}$ -- Binned with $j=15$}
\end{subfigure}

\begin{subfigure}{0.49\textwidth}
    \centering 
\includegraphics[width=1\textwidth]{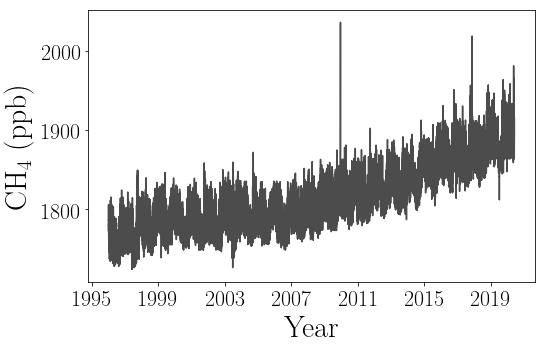}
\caption{Mauna Loa $CH_4$ -- Raw}
\end{subfigure}
\begin{subfigure}{0.49\textwidth}
    \centering 
\includegraphics[width=1\textwidth]{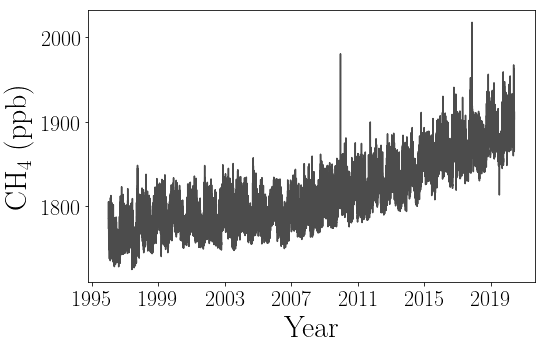}
\caption{Mauna Loa $CH_4$ -- Binned with $j=4$}
\end{subfigure}

\vspace*{3mm}
\caption{\textbf{Data sets}}
\label{fig:mix-supp}
\end{figure}

\begin{figure}[h]
    \centering
\includegraphics[width=0.45\textwidth]{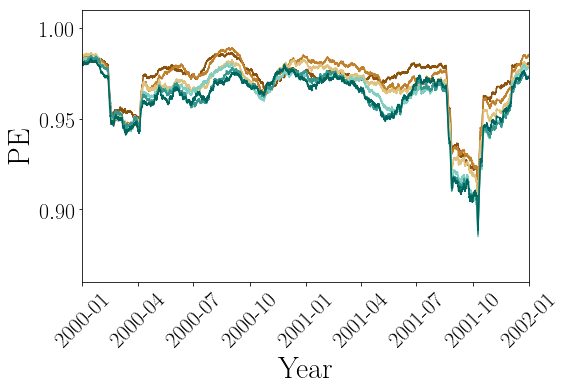}
\includegraphics[width=0.45\textwidth]{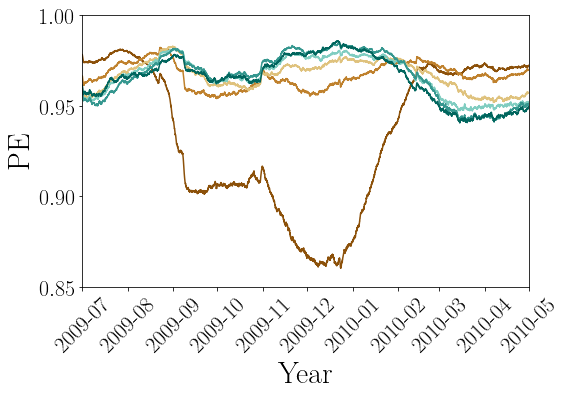}
\caption{\textbf{Mauna Loa data:} Close-ups showing PE traces of the raw Mauna Loa data.}
\label{fig:ml-closeup}
\end{figure}

\end{document}